\documentclass[11pt,tbtags,reqno,A4]{amsart}

\usepackage{amsmath,amsthm,url}

\usepackage{pstricks,graphicx}

\numberwithin{equation}{section}

\newcommand{\ima}{\ensuremath{\mathrm{i}}}
\newcommand{\tr}{\ensuremath{\,\mathrm{tr}\,}}
\newcommand{\trans}{\ensuremath{\mathrm{t}}}
\newcommand{\X}{\ensuremath{\mathbf{X}}}
\newcommand{\U}{\ensuremath{\mathbf{U}}}
\newcommand{\Y}{\ensuremath{\mathbf{Y}}}
\newcommand{\bS}{\ensuremath{\mathbf{S}}}
\newcommand{\W}{\ensuremath{\mathbf{W}}}
\newcommand{\F}{\ensuremath{\mathbf{F}}}
\newcommand{\I}{\ensuremath{\mathbf{1}}}
\newcommand{\bO}{\ensuremath{\mathbf{0}}}

\newcommand{\LL}{\ensuremath{\langle\!\langle}}
\newcommand{\RR}{\ensuremath{\rangle\!\rangle}}
\newcommand{\LLb}{\ensuremath{{\Big\langle}\!\!{\Big\langle}}}
\newcommand{\RRb}{\ensuremath{{\Big\rangle}\!\!{\Big\rangle}}}

\newcommand{\GbE}{\ensuremath{\mathrm{G}\:\!\!\beta\!\!\;\mathrm{E}}}
\newcommand{\GBE}{\ensuremath{\mathrm{G}\:\!\!\beta'\!\!\;\mathrm{E}}}
\newcommand{\LbE}{\ensuremath{\mathrm{ch}\:\!\!\beta\!\!\;\mathrm{E}}}
\newcommand{\LBE}{\ensuremath{\mathrm{ch}\:\!\!\beta'\!\!\;\mathrm{E}}}

\newtheorem{theorem}{Theorem}
\newtheorem{lemma}[theorem]{Lemma}
\newtheorem{proposition}[theorem]{Proposition}

\theoremstyle{remark}

\newtheorem*{acknow}{Acknowledgments}

\def\Square{\pspicture(0.5,0.5)\psframe[dimen=middle](0.5,0.5)\endpspicture}

\begin{document}

\title[Duality in random matrix ensembles]{Duality in random matrix ensembles for all $\beta$}

\author{Patrick Desrosiers}

\address{Institut de Physique Th\'eorique, CEA--Saclay, 91191
 Gif-sur-Yvette cedex, France.}
\email{patrick.desrosiers@cea.fr}

\thanks{IPhT--t08/013}

\begin{abstract}
 Gaussian  and Chiral $\beta$-Ensembles, which generalise well known orthogonal ($\beta=1$), unitary ($\beta=2$), and symplectic ($\beta=4$) ensembles of random Hermitian  matrices, are considered.  Averages are shown to satisfy duality relations like $\{\beta,N,n\}\Leftrightarrow \{4/\beta,n,N\}$ for all $\beta>0$, where $N$ and $n$ respectively denote the  number of eigenvalues and  products of characteristic polynomials.    At the edge of the spectrum, matrix integrals of the Airy (Kontsevich) type are obtained.  Consequences on the integral representation of the multiple orthogonal polynomials and the partition function of the formal one-matrix model are also discussed.  Proofs rely on the theory of multivariate symmetric polynomials, especially Jack polynomials.
\end{abstract}



\maketitle

{

\renewcommand{\baselinestretch}{1.0}
\setlength{\parskip}{0ex}
\small
\tableofcontents

}

\newpage

\renewcommand{\baselinestretch}{1.0}
\setlength{\parskip}{1ex}

\normalsize

\section{Introduction}

It is the purpose of this article to obtain new dualities between different ensembles of random variables inspired by matrix models.  Simple duality relations are given below, after a short review of the $\beta$-Ensembles.

\subsection{$\beta$-Ensembles}

Let $x=(x_1,\ldots,x_N)$ denote a set of $N$ random variables.  Moreover, let $\beta$  be a non-negative real number; it is called the Dyson index.  The joint probability density function (p.d.f.) for the Gaussian $\beta$-Ensemble ($\GbE$) is
\begin{equation}\label{gbe}
 P_{{ \GbE_N}}(x)dx=\frac{1}{Z_{\GbE_N}}\prod_{1\leq i<j\leq N}|x_i-x_j|^\beta \prod_{i=1}^N e^{-x_i^2}dx_i.
\end{equation}
The average of a function $f$ over the ensemble is given  by
\begin{equation}
 \Big\langle f(x) \Big\rangle_{x\in \GbE_N}:= \int_{\mathbb{R}^N} f(x)P_{\GbE_N}(x)dx.
\end{equation}
Note that $Z_{\GbE_N}$ is chosen such that the average of the identity equals one.  Physically, the density $P_{\GbE_N}$ can be interpreted as the Boltzmann factor of a classical (two-dimensional) Coulomb gas at temperature $1/\beta$ \cite{DysonI}.  There is also a quantum mechanical interpretation of the density: $\sqrt{P_{\GbE_N}}$ is  the ground state of a $N$-body problem with pairwise potential interaction of the form $1/r^2$ \cite{Calogero}. The chiral Gaussian $\beta$-Ensemble ($\LbE$) is defined similarly:
\begin{equation}\label{lbe}
 P_{\LbE^\gamma_N}(x)dx=\frac{1}{Z_{\LbE^\gamma_N}}\prod_{1\leq i<j\leq N}|x_i-x_j|^\beta \prod_{i=1}^N x_i^\gamma e^{-x_i}dx_i,
\end{equation}
where the variables $x_i$ are positive reals and the real part of $\gamma$ is greater then $-1$.  Note that the Gaussian and chiral Gaussian ensembles are sometimes called Hermite and Laguerre (or even Wishart) ensembles respectively (cf.\ \cite{DF,Dumitriu}).

In this article, the $\beta$-Ensembles are considered from a Random Matrix Theory \cite{ForresterBook,Mehta} perspective.  Let $\mathbf{X}=[X_{i,j}]$ be a random $N\times N$  Hermitian matrix whose entries are real ($\beta=1$), complex ($\beta=2$), or quaternion real ($\beta=4$) (see Appendix A for more detail).  The trivial $\beta=0$ case corresponds to a real diagonal matrix.   It is a classical result that Eq.~\eqref{gbe} provides the joint density for the eigenvalues of a Hermitian matrix  $\mathbf{X}$ when the latter is drawn with probability
\begin{equation}\label{Mgbe}
 e^{-\tr \mathbf{X}^2}(d\mathbf{X}), \quad
\end{equation}
where $(d\mathbf{X})$ stands for the normalised product of all the real independent elements of $[dX_{i,j}]$.  Similarly, if a positive definite Hermitian matrix $\mathbf{X}$ is distributed according to
\begin{equation}\label{Mlbe}
 (\det \mathbf{X})^\gamma e^{-\tr \mathbf{X}}(d\mathbf{X}),
\end{equation}
then the p.d.f.\ of its eigenvalues is given by Eq.~\eqref{lbe}.  Moreover, set $\mathbf{X}=\mathbf{Y}^\dagger\mathbf{Y}$, where $\mathbf{Y}$ is a $N_1\times N_2$ rectangular matrix such that $N_1\geq  N_2$.  One can show that if $\mathbf{Y}$ has a Gaussian distribution, then the p.d.f.\ of $\mathbf{X}$  is given by \eqref{Mlbe} with $\gamma=(\beta/2)(N_1-N_2+1-2/\beta)$ \cite[Chapter I]{ForresterBook}.

It is worth mentioning that Eqs.\ \eqref{gbe} and \eqref{lbe} can also be realised, for all $\beta>0$,  as the eigenvalue p.d.f.\  of tri-diagonal real symmetric matrices \cite{Dumitriu}.  In the large $N$ limit, scaled versions of these tri-diagonal matrices can be seen as stochastic differential operators \cite{EdelmanSutton, Ramirez}.

\subsection{Some results}

The aim of this paper is to prove the equivalence of some averages over $\beta$-ensembles and other averages over $4/\beta$-ensembles  The averages contain a source matrix $\mathbf{S}$  of size $n$,  and an external field matrix $\mathbf{F}$ of size $N$.  These matrices are Hermitian with eigenvalues $s=(s_1,\ldots,s_N)$ and $f=(x_1,\ldots,x_N)$ respectively.  Roughly speaking, the duality relations obtained here read
\begin{equation}\label{duality}
\left\lbrace
\begin{array}{c}
\,\,\,\beta\,\,\,\\
N\\
\F\\
n\\
 \bS
\end{array}
\right\rbrace
\Longleftrightarrow
 \left\lbrace
\begin{array}{c}
4/\beta\\
n\\
\bS\\
N\\
 \F
\end{array}
\right\rbrace.
\end{equation}

In order to formulate the results precisely, further notation has to be introduced.  Let $\mathbf{1}$ be the identity matrix and let
 \begin{equation}
 \left\langle F(\mathbf{X})\right\rangle_{\mathbf{X}\in\GbE}=\int (d\mathbf{X})e^{-\tr\mathbf{X}^2} F(\mathbf{X})\Big/ \int (d\mathbf{X})e^{-\tr\mathbf{X}^2},
 \end{equation}
 which is the average of the complex-valued function $F$ over the Gaussian $\beta$-Ensemble of matrices with p.d.f.\ \eqref{Mgbe} (not to be confused with the average over the eigenvalues).  Special realisations of Eq.\ \eqref{duality} are given in the following propositions.  Only expression relative to the Gaussian ensembles will be displayed for the moment; dualities for the Chiral ensembles will be given in Sections 3 and 4.

\begin{proposition}\label{prop1}
Suppose that $\beta=1,2$ or $4$ and that $\beta'=4/\beta$.  Then
\begin{multline}\label{DualMat1}
e^{-\mathrm{tr}\F^2}\left\langle\prod_{j=1}^n\det\left(s_j\I\pm\mathrm{i}\sqrt{\frac{2}{\beta}}\X\right)e^{2\tr\X\F}\right\rangle_{\X\in\GbE_N}=\\
e^{-\mathrm{tr}\bS^2}\left\langle\prod_{j=1}^N\det\left(\Y\pm\mathrm{i}\sqrt{\frac{2}{\beta}}f_j\I\right)e^{2\tr \Y\bS}\right\rangle_{\Y\in \GBE_n}.
\end{multline}
\end{proposition}
The previous result concerns the average of products of characteristic polynomials.  Special cases of this duality previously appeared in the literature.  For instance,  the duality for moments of characteristic polynomials, which corresponds to $s_1=\ldots=s_n$ and $\F=\bO$,  has been proved for $\beta=1,2,4$ and conjectured for all $\beta$ by Mehta and Normand \cite{Normand}.  For $\F=\bO$ (i.e., no external field) and $\beta=2$, but for distinct $s$, this has been observed by Fyodorov and Strahov in \cite{FyodorovStrahaov}.   Based on a previous work \cite{BrezinHikami00} and   the supersymmetric method, Br\'ezin and Hikami \cite{BrezinHikami01} have obtained   dualities reproducing Eq.\ \eqref{DualMat1} in the case $\F=\bO$ and $\beta=1,2,4$.  The problem with a non-zero external field and $\beta=2$,  has been recently solved by same authors \cite{BrezinHikami07}.

A less common duality of the type \eqref{duality}, but not affecting $\beta$,  concerns  products of inverse characteristic polynomials.

\begin{proposition}\label{prop2}Suppose that $\beta=1,2$ or $4$ and that $\beta'=4/\beta$.  Assume moreover that the variables $s$  and $f$ have non-zero imaginary parts. Then
\begin{multline}
e^{-\tr \F^2}\left\langle\prod_{j=1}^n\det(s_j\I \pm  \X)^{-\beta/2}e^{2\tr\X\F}\right\rangle_{\X\in\GbE_N}=\\
e^{-\tr\bS^2}\left\langle\prod_{j=1}^N\det(\Y\pm f_j\I)^{-\beta/2}e^{2\tr\Y\bS}\right\rangle_{\Y\in\GbE_n}.
\end{multline}
\end{proposition}

More general dualities will be exposed in Sections 3 and 4 after the introduction of multivariate  hypergeometric functions in Section 2.  To the best of the author's knowledge,  explicit dualities valid for  $\beta$ general and $N$ finite,  first appeared in a 1997  paper by Baker and Forrester \cite{Baker}.  For instance, Eq.\ 5.31 in \cite{Baker} is equivalent to Eq.\ \eqref{DualMat1} above with $\F=\bO$ and $s_1=\ldots=s_n=t$ say, and $\beta$ integer.  The latter expression has been used in \cite{DF} for calculating asymptotic corrections to the global eigenvalue density when $\beta$ is even.
The present work can be considered, to some extend,  as a continuation of \cite{Baker}.   Note also that Proposition~7, for $s_1=\ldots=s_n=t$ and $f=0$,  provides a proof of the formula conjectured by Mehta and Normand  for all $\beta>0$ \cite[Eq. 3.29]{Normand}.

Closely related are the ``particle-hole'' dualities observed, when $\beta$ is rational,  in the limit  $N\rightarrow\infty$ of the correlation functions for the Sutherland model or equivalently, for the Circular $\beta$-Ensembles (see for instance \cite{Serban} and references therein).  This can also be interpreted as a strong-weak coupling duality (see for example \cite{Jonke}).

Inter-relationships between orthogonal, unitary and symplectic ensembles have been also considered, at the level of the joint eigenvalue p.d.f.\ itself, for eigenvalues respecting interlacing inequalities \cite{ForresterRains}.  Very recently, this work has been generalised by Forrester in \cite{Forrester07}.  The duality found in the latter reference can be summarised as follows : setting $\beta'=2(r+1)=4/\beta$, the joint distribution of every $(r+1)$-st eigenvalue in a $\beta$-ensemble is equal to that of a $\beta'$-ensemble if the eigenvalues are properly ordered.

Before going further into the study of dualities, a last comment is in order.  For $\beta=1,2$ and $4$, expectation values of products and ratios of characteristic polynomials  can be expressed in terms of determinants or Pfaffians (see the extensive study by Borodin and Strahov \cite{BorodinStahov} and references therein).  The size of the determinants depends on the number $n$ of characteristic polynomials  but not on the size $N$ of the random matrix.  Thus, one can exploit the determinant formulae for calculating the asymptotic behaviour of the correlation functions when $N\rightarrow\infty$.  For general $\beta$, no such determinantal formulae are available.  Then, solving a $\beta$-matrix model amounts to finding a reduced integral representation for the correlation functions.  In other words, the aim is to obtain integral formulae whose dimension does not depend on the size of the matrix, thus providing a representation for the correlation functions that, in theory, allows to take the limit $N\rightarrow\infty$.

\section{Preliminary definitions}
 This section furnishes a brief introduction to the theory of symmetric polynomials.  More detail can be found in \cite{ForresterBook,Kadell,Mac,Stan}

\subsection{Partitions}
Let $\lambda=(\lambda_1,\lambda_2,\ldots,)$ denote a partition of $n$; that is, a sequence of non-negative integers such that
\begin{equation}
 \lambda_1\geq\lambda_2\geq \ldots\geq0,\qquad |\lambda|:=\sum_i\lambda_i=n.
\end{equation}
One usually writes $(\lambda_1,\ldots,\lambda_\ell,0,\ldots,0)=(\lambda_1,\ldots,\lambda_\ell)$, where $\ell=\ell(\lambda)$ gives the number of non-zero parts in $\lambda$.  A partition can also be expressed as follows: $\lambda=(\ldots,3^{n_3}, 2^{n_2},1^{n_1})$  where $n_k=n_k(\lambda)$ is the number of parts of $\lambda$ that are equal to $k$.  The conjugate partition of $\lambda$, written $\lambda'=(\lambda'_1,\lambda'_2,\ldots)$, is such that
\begin{equation}
 \lambda'_k=\#\{\lambda_i \in \lambda : \lambda_i\geq k\}
\end{equation}

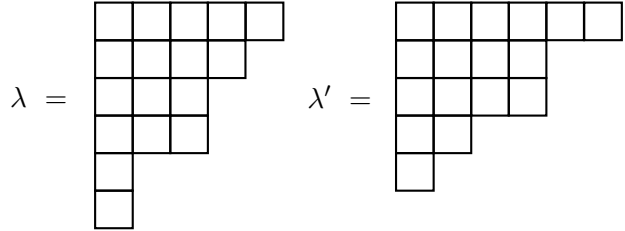
\begin{figure}[h]\caption{Diagram and conjugate diagram}\label{fig1}
\begin{center}
 \begin{pspicture}(0,1)(9,6)
 {
\rput(0.5,4){$\lambda\,\,=$}
{

\rput(1.5,5){\Square}\rput(2,5){\Square}\rput(2.5,5){\Square}\rput(3,5){\Square}\rput(3.5,5){\Square}
\rput(1.5,4.5){\Square}\rput(2,4.5){\Square}\rput(2.5,4.5){\Square}\rput(3,4.5){\Square}
\rput(1.5,4){\Square}\rput(2,4){\Square}\rput(2.5,4){\Square}
\rput(1.5,3.5){\Square}\rput(2,3.5){\Square}\rput(2.5,3.5){\Square}
\rput(1.5,3.0){\Square}
\rput(1.5,2.5){\Square}

     }

\rput(4.5,4){$\lambda'\,\,=$}
{

\rput(5.5,5){\Square}\rput(6,5){\Square}\rput(6.5,5){\Square}\rput(7,5){\Square}\rput(7.5,5){\Square}\rput(8,5){\Square}
\rput(5.5,4.5){\Square}\rput(6,4.5){\Square}\rput(6.5,4.5){\Square}\rput(7,4.5){\Square}
\rput(5.5,4){\Square}\rput(6,4){\Square}\rput(6.5,4){\Square}\rput(7,4){\Square}
\rput(5.5,3.5){\Square}\rput(6,3.5){\Square}
\rput(5.5,3.0){\Square}
     }

}
 \end{pspicture}
\end{center}
\end{figure}

\begin{figure}[h]\caption{Arms and legs of $s=(i,j)\in\lambda$  }\label{fig2}
\begin{center}
 \begin{pspicture}(-1,0)(5,4.5)
 {

\psline{-}(0,0)(0,4)(6,4)(6,3)(5,3)(5,1)(3,1)(3,0.5)(1,0.5)(1,0)(0,0)

\psline[linestyle=dotted]{<->}(0,2)(1.75,2)
\rput(1,2.25){\footnotesize $a'_\lambda(s)$}

\psline[linestyle=dotted]{<->}(2.25,2)(5,2)
\rput(3.75,2.25){\footnotesize $a_\lambda(s)$}

\psline[linestyle=dotted]{<->}(2,0.5)(2,1.75)
\rput(2.5,1.2){\footnotesize $l_\lambda(s)$}

\psline[linestyle=dotted]{<->}(2,2.25)(2,4)
\rput(2.5,3.0){\footnotesize $l'_\lambda(s)$}

\rput(2,2){\Square}
\rput(2,2){\footnotesize$s$}

\rput(-0.75,2){$\lambda\;=$}

}\end{pspicture}
\end{center}
\end{figure}
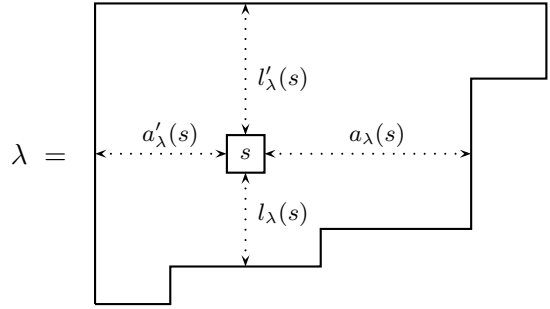

To each partition, we associate a diagram by drawing $\lambda_1$ boxes on the first row, then $\lambda_2$ boxes under the first row, and so on, all boxes left justified. See Fig.\ \ref{fig1}.   The conjugation of the partition then corresponds to the transposition (as for matrices) of the diagram.   The arm of a point $s=(i,j)$ in $\lambda$, written $a_\lambda(s)$, is the number of boxes to the right of $(i,j)$ in the $i$th row of the diagram $\lambda$.  As illustrated in Fig.\ \ref{fig2}, similar definitions exist for the leg $l$,  co-arm $a'$, and  co-leg $l'$ of the point $s=(i,j)$.

The dominance ordering of partitions is defined as follows:
\begin{equation}
 \lambda\geq\mu \qquad\Longleftrightarrow\qquad  \sum_{i=1}^k(\lambda_i-\mu_i)\geq0\qquad \forall \, k.
\end{equation}
This ordering is partial.  Note the obvious property:
\begin{equation}
 \lambda\geq\mu\qquad\Longleftrightarrow\qquad \lambda'\leq\mu'.
\end{equation}

Four functions on partitions will be frequently used in the article.  They are:
the generalised factorial
\begin{equation}
 [u]^{(\alpha)}_\lambda:=\prod_{j\geq1}\frac{\Gamma\left(u-(j-1)/\alpha+\lambda_j\right)}{\Gamma\left(u-(j-1)/\alpha \right)};
\end{equation}
the specialisation coefficient
\begin{equation}
 b^{(\alpha,N)}_\lambda=\prod_{s\in\lambda}\left(N-l'_\lambda(s)+\alpha a'_\lambda(s)\right);
\end{equation}
and the ``lower'' and ``upper'' hook-lengths \cite{Stan} \footnote{$h^\lambda_{(\alpha)}$ and $h_\lambda^{(\alpha)}$ are respectively noted  $h_\lambda$ and $d'_\lambda$ in \cite{ForresterBook},  while they are noted $c_\lambda$ and $c'_\lambda$ in \cite{Mac}.}
\begin{equation}\label{eqHook}
 h^\lambda_{(\alpha)}=\prod_{s\in\lambda}\left(l_\lambda(s)+1+\alpha a_\lambda(s)\right)\quad\mbox{and}\quad  h_\lambda^{(\alpha)}=\prod_{s\in\lambda}\left(l_\lambda(s)+\alpha+\alpha\ a_\lambda(s)\right);\
\end{equation}
Note that $b^{(\alpha,N)}_\lambda=\alpha^{|\lambda|}[N/\alpha]^{(\alpha,N)}_\lambda$.

\subsection{Jack polynomials}

Let $x$ stand for the ordered set $(x_1,\ldots,x_N)$.  A function $f$ is symmetric if it is invariant under permutations of the variables; that is, $f(x)=f(x_\sigma )$ for any element  $\sigma$ of the symmetric group $S_N$, where $x_\sigma =(x_{\sigma(1)},\ldots,x_{\sigma(N)})$.  The set of all symmetric polynomials in $N$ variables, whose coefficients are rational functions of $\alpha$,  form an algebra over the ring $\mathbb{Q}(\alpha)$, witten $P^{S_N}$.  A symmetric polynomial that is homogeneous of degree $n$ can be decomposed into the monomial basis $\{m_\lambda : |\lambda|=n\}$, where
\begin{equation}
 m_\lambda(x)=\frac{1}{n(\lambda)!}\sum_{\sigma\in S_N}x_\sigma^\lambda=\frac{1}{n(\lambda)!}\sum_{\sigma\in S_N}x_{\sigma(1)}^{\lambda_1}\cdots x_{\sigma(N)}^{\lambda_N}
\end{equation}
and
$
 n(\lambda)!:=n_1(\lambda)! n_2(\lambda)!\cdots
$.

Another important basis for $P^{S_N}$ is provided by products of power sums $\{p_\lambda\}$, where
\begin{equation}
 p_\lambda(x):=p_{\lambda_1}(x)\cdots p_{\lambda_\ell}(x),\qquad p_k(x)=\sum_{i=1}^Nx_i^k.
\end{equation}
The combinatorial (or Fock space) scalar product in the algebra of symmetric polynomials can be defined by
\begin{equation}
 \label{CombinatScalProd}
\LL p_\lambda|p_\mu\RR^{(\alpha)}=\alpha^{\ell(\lambda)}z_\lambda\delta_{\lambda,\mu},
\end{equation}
where  $z_\lambda=\prod_k k^{n_k}k!$. The parameter $\alpha$ is related to Dyson's  $\beta$ index as follows:
\begin{equation}\label{alphabeta}
 \alpha=\frac{2}{\beta}.
\end{equation}

 The Macdonald automorphism will play an important role in the following paragraphs; it is given by
\begin{equation}\label{defHomo}
 \omega_k p_n=(-1)^{n-1}k\,p_n
\end{equation}
 and satisfies
\begin{equation}\label{EqHomo}
\LL\omega_k f|g\RR^{(\alpha)}=\LL f|\omega_kg\RR^{(\alpha)},\qquad
 \LL\omega_{1/\alpha} f|g\RR^{(\alpha)}=\LL \omega_1 f|g\RR^{(1)}.
\end{equation}
for any symmetric polynomials $f$ and $g$.

The (monic) Jack polynomials, denoted by $P_\lambda(x)=P_\lambda(x;\alpha)$ or by $P^{(\alpha)}_\lambda(x)$, generalise many important symmetric polynomials:
\begin{equation}
 P^{(\alpha)}_\lambda=\begin{cases}e_{\lambda'},& \alpha=0\\
s_\lambda, &\alpha=1\\
Z_\lambda/h^\lambda_{(\alpha)}, &\alpha=2\\
m_\lambda&\alpha=\infty
\end{cases}
\end{equation}
where $e_\lambda$, $s_\lambda$, and $Z_\lambda$ respectively stand for the elementary,  Schur, and Zonal polynomials.
$P^{(\alpha)}_\lambda$
is the unique symmetric polynomial  with coefficients in $\mathbb{Q}(\alpha)$ satisfying
\begin{equation}\label{eqOrthoJack}
 \begin{array}{lll}
 (1)&\displaystyle P_{\lambda} = m_{\lambda} +\sum_{\mu < \lambda} c_{\lambda \mu}(\alpha) m_{\lambda}&\mbox{(triangularity)}\\
(2)&\displaystyle  \LL P_{\lambda}| P_{\mu} \RR^{(\alpha)} =\Vert P_{\mu}^{(\alpha)}\Vert^2 \delta_{\lambda,\mu} &\mbox{(orthogonality)}
\end{array}
\end{equation}
Alternatively, the Jack polynomials can be considered as the unique triangular symmetric polynomials that comply with
\begin{equation}
 D\,P_\lambda(x)=e_\lambda(\alpha) P_\lambda (x)\qquad \mbox{(eigenfunctions)},
\end{equation}
where
\begin{equation}
 D=\sum_{i=1}^N\left(\alpha x_i^2\partial_{x_i}^2+2\sum_{j\neq i}\frac{x_ix_j}{x_i-x_j}\partial_{x_i}\right).
\end{equation}
The eigenvalues are given by  $e_\lambda(\alpha)=\sum_{i\geq1}\left(\alpha(i-1)\lambda'_i-(i-1)\lambda_i\right)$. The following specialisation and normalisation formulae will be used later:
\begin{equation}\label{eqSpecial}
 P^{(\alpha)}_\lambda (1^N)=\frac{b_\lambda^{(\alpha,N)}}{h^\lambda_{(\alpha)}}\quad\mbox{and}\quad \Vert P^{(\alpha)}_\lambda\Vert^2=\frac{h_\lambda^{(\alpha)}}{h^\lambda_{(\alpha)}}.
\end{equation}

The Jack polynomials possess several remarkable properties.  Amongst them, it is worth mentioning their duality (see eq.\  \eqref{EqHomo})
\begin{equation}\label{eqDualityJack}
 \omega_\alpha \, P^{(\alpha)}_\lambda =\frac{1}{ \Vert P^{(\alpha')}_{\lambda'}\Vert^2} P_{\lambda'}^{(\alpha')}\qquad\mbox{with}\qquad \alpha':=\frac{1}{\alpha},
\end{equation}
and the Cauchy type formula
\begin{equation}\label{eqCauchy}
 \prod_{i=1}^N\prod_{j=1}^M\frac{1}{(1-x_iy_j)^{1/\alpha}}=\sum_{\lambda}\frac{1}{\Vert P^{(\alpha)}_\lambda\Vert^2}P^{(\alpha)}_\lambda (x)\,P^{(\alpha)}_\lambda (y).
\end{equation}
By applying the former property to the latter equation, one gets
\begin{equation}
 \prod_{i=1}^N\prod_{j=1}^M(1+x_iy_j)=\sum_{\lambda}P^{(\alpha)}_\lambda (x)\,P^{(\alpha')}_{\lambda'} (y).
\end{equation}

\subsection{Multivariate hypergeometric functions}

In multivariate analysis, the Jack polynomial $P_{\lambda}(x_1,x_2,\ldots)$ plays a role that is  similar to that played by the  monomial $x_1^{|\lambda|}$ for functions in one variable.  As an example, the multivariate hypergeometric functions of two sets of variables are given by \cite{ForresterBook, Yan}:
\begin{multline}\label{eqHyper}
 {\,_p{\mathcal{F}}_q}^{(\alpha)} (a_1,\ldots,a_p;b_1,\ldots,b_q;x_1,\ldots,x_N; y_1,\ldots,y_N)=\\
\sum_{\lambda}\frac{\alpha^{|\lambda|}}{h_\lambda^{(\alpha)}}\frac{[a_1]^{(\alpha)}_\lambda\cdots[a_p]^{(\alpha)}_\lambda}{[b_1]^{(\alpha)}_\lambda\cdots [b_q]^{(\alpha)}_\lambda}\frac{P_\lambda^{(\alpha)}(x) P_\lambda^{(\alpha)}(y)}{P^{(\alpha)}_\lambda(1^N)}.
\end{multline}
There is no simple ``explicit forms'' for the latter functions in general. However, one can prove that \cite{Kaneko}
\begin{equation}\label{eqHyperspecial}
  {\,_0{\mathcal{F}}_0}^{(\alpha)}(x_1,\ldots,x_N;t,\ldots,t)=\prod_{i=1}^Ne^{tx_i}
\end{equation}
\begin{equation}
 {\,_1{\mathcal{F}}_0}^{(\alpha)}(a;x_1,\ldots,x_N;t,\ldots,t)=\prod_{i=1}^N(1-tx_i)^{-a}.
\end{equation}
Softwares are also available for computing ``truncated'' hypergeometric functions; they are based on the exact calculation of Jack polynomials (e.g., see \cite{Koev}).

As explained in \cite{Baker}, the ${\,_p{\mathcal{F}}_q}^{(\alpha)}$'s  provide  generating functions for the multivariate Hermite  and Laguerre polynomials, respectively written $\mathcal{H}_\lambda^{(\alpha)}$ and $\mathcal{L}_\lambda^{(\alpha,\gamma)}$.  These polynomials were first defined by Lassalle in \cite{ LassalleLaguerre,LassalleHermite}.   Explicitly, \footnote{All polynomials involved in Eqs.\ \eqref{eqHermite} and \eqref{eqLaguerre} are monic while those in \cite{Baker} are not.  One goes from a convention to another via $\mathcal{H}_\lambda^{(\alpha)}(x)=2^{-|\lambda|}{P}_\lambda^{(\alpha)}(1^N) {H}_\lambda (x;\alpha)$ and $\mathcal{L}_\lambda^{(\alpha,\gamma)}(x)=(-1)^{|\lambda|}|\lambda|!{P}_\lambda^{(\alpha)}(1^N) {L}_\lambda^{\gamma}(x;\alpha)$.  Additionally, ${P}_\lambda^{(\alpha)}(x)=\alpha^{-|\lambda|}h_\lambda^{(\alpha)} C^{(\alpha)}_\lambda (x)/|\lambda|!$.}
\begin{equation}\label{eqHermite}
 e^{-p_2(y)}{\,_0{\mathcal{F}}_0}^{(\alpha)}(2x;y)=\sum_\lambda \frac{1}{A_\lambda(\alpha,N)} \mathcal{H}_\lambda^{(\alpha)}(x){P}_\lambda^{(\alpha)}(y),
\end{equation}
where
\begin{equation}
 A_\lambda(\alpha,N)=\frac{1}{(2\alpha)^{|\lambda|}}h_\lambda^{(\alpha)}{P}_\lambda^{(\alpha)}(1^N),
\end{equation} and
\begin{equation}\label{eqLaguerre}
 e^{-p_1(y)}{\,_0{\mathcal{F}}_1}^{(\alpha)}(\gamma+q;x;y)=\sum_\lambda \frac{1}{B_\lambda(\alpha,\gamma,N)} \mathcal{L}_\lambda^{(\alpha,\gamma)}(x){P}_\lambda^{(\alpha)}(y),
\end{equation}
where
\begin{equation}
  B_\lambda(\alpha,\gamma,N)=2^{|\lambda|}[\gamma+q]^{(\alpha)}_\lambda A_\lambda(\alpha,N), \qquad q=1+(N-1)/\alpha.
\end{equation}
Multivariate orthogonal polynomials are deeply connected to the Dunkl operators and Calogero systems \cite{Baker, vDiejen,Rosler}.  From these relations, it is possible to show,  for instance, that the Hermite and Laguerre polynomials in many variables provide orthogonal bases for the algebra of symmetric polynomials:
 \begin{equation}\label{eqOrthoH}
 \Big\langle \mathcal{H}_\lambda^{(2/\beta)}(x) \mathcal{H}_\mu^{(2/\beta)}(x) \Big\rangle_{x\in \GbE_N}=A_\lambda(2/\beta,N)\delta_{\lambda,\mu}
\end{equation}
\begin{equation}\label{eqOrthoL}
\Big\langle \mathcal{L}_\lambda^{(2/\beta,\gamma)}(x) \mathcal{L}_\mu^{(2/\beta,\gamma)}(x) \Big\rangle_{x\in \LbE^\gamma_N}=B_\lambda(2/\beta,\gamma,N)\delta_{\lambda,\mu}.
\end{equation}

\section{Expectation values of Jack polynomials}

In this section, it is proved that the averages of Jack polynomials over the $\beta$-ensembles defined in \eqref{gbe} and \eqref{lbe}  enjoy simple duality properties.  Recall that the  Jack polynomials form a basis for symmetric polynomials.  Thus, in principle at least, dualities involving Jack polynomials provide  general tools for reducing the dimension of the correlation functions.

\subsection{Gaussian ensembles}

The starting point is a conjecture by Goulden and Jackson \cite{Goulden} first proved by Okounkov in \cite{Okounkov}.  Another proof is given below.

\begin{lemma}\label{lemma1}
Let $\lambda$ be a partition of an even integer and let $\mu$ be the partition $(2^{|\lambda|/2})$.  Also, let $\LL f|g\RR$ denote the combinatorial scalar product denied in Eq.\ \eqref{CombinatScalProd}.  Then
\begin{equation}\label{EqProof1}
\left\langle {P^{(2/\beta)}_\lambda(x)}\right\rangle_{x\in\GbE_N}=\left(\frac{\beta}{4}\right)^{\ell(\mu)}\frac{b_\lambda^{(2/\beta,N)}}{\ell(\mu)!}\LLb
 P^{(2/\beta)}_\Lambda\,\Big|\,p_{\mu}\RRb^{(2/\beta)}.
\end{equation}
Alternatively,
\begin{equation}\label{EqProof1}
\left\langle {P^{(2/\beta)}_\lambda(x)}\right\rangle_{x\in\GbE_N}={b_\lambda^{(2/\beta,N)}}\underset{p_\mu}{\mathrm{coeff}}
 P^{(2/\beta)}_\Lambda,
\end{equation}
where $\underset{x}{\mathrm{coeff}}f$ denote the coefficient of $x$ in  $f$.
\end{lemma}
\begin{proof}It is simpler to set $\alpha=2/\beta$.  The orthogonality \eqref{eqOrthoH} and Eq.\ \eqref{eqHermite} immediately imply \cite[Proposition 3.8]{Baker}
\begin{equation}
 \left\langle{\,_0{\mathcal{F}}_0}^{(\alpha)}(2x;y){\,_0{\mathcal{F}}_0}^{(\alpha)}(2x;z) \right\rangle_{x\in\GbE}=e^{p_2(y)+p_2(z)}{\,_0{\mathcal{F}}_0}^{(\alpha)}(2y;z).
\end{equation}
Then, setting $z=0$ in the last equation and recalling that $P_\lambda(0)$ is equal to $1$  if $\lambda=(0)$ and 0 otherwise, one gets
\begin{equation}
 \left\langle{\,_0{\mathcal{F}}_0}^{(\alpha)}(2x;y) \right\rangle_{x\in\GbE}=e^{p_2(y)}.
\end{equation}
Now, the left-hand-side is developted in terms of the Jack polynomials and the combinatorial scalar product with respect to $P_\lambda^{(\alpha)}$ is taken (see Eq.\ \eqref{eqOrthoJack}).  This leads to
\begin{equation}
 \frac{(2\alpha)^{|\lambda|/2}}{h^\lambda_{(\alpha)}P^{(\alpha)}_\lambda(1^N)}\left\langle P^{(\alpha)}_\lambda(x)\right\rangle_{x\in\GbE}= \frac{1}{(|\lambda|/2)!}\LL  P^{(\alpha)}_\lambda| (p_2)^{|\lambda|/2}\RR^{(\alpha)},
\end{equation}
which, by virtue of Eq.\eqref{eqSpecial}, is equivalent to the first equation of the lemma.  The second equation simply follows from
\begin{equation}
 \LL p_\mu|p_\mu \RR^{(\alpha)} = (2\alpha)^{|\lambda|/2}({|\lambda|}/{2})!\quad \mbox{if}\quad \mu=(2^{|\lambda|/2}).
\end{equation}

\end{proof}

\begin{proposition} \label{theo1}
Set $x'=\mathrm{i}x\sqrt{2/{\beta}}$ and ${\beta'}=4/\beta>0$.
  Then, for every $N\geq\ell(\lambda)$ and $N'\geq \ell(\lambda')$,
\begin{equation}\label{eqTheo1}
\left\langle \frac{P^{(2/\beta)}_\lambda(x)}{P^{(2/\beta)}_\lambda(1^{N})}\right\rangle_{x\in\GbE_N}=\left\langle
 \frac{P^{(2/\beta')}_{\lambda'}(x')}{P^{(2/\beta')}_{\lambda'}(1^{N'})}\right\rangle_{x\in\GBE_{N'}}.
\end{equation}
Note that, as a consequence of the orthogonality \eqref{eqOrthoH}, the expectation values are non-zero only
 if $|\lambda|$ is even.
\end{proposition}
\begin{proof}
Assume $N\geq \ell(\lambda)$ and  $|\lambda|$ even.  Following the notation used for the previous lemma, one has
\begin{equation}\label{eqProofTheo1}
 \left\langle \frac{P^{(\alpha)}_\lambda(x)}{P^{(\alpha)}_\lambda(1^N)}\right\rangle_{x\in\GbE_N}=\left(\frac{2}{\alpha}\right)^{\ell(\mu)}\frac{h^\lambda_{(\alpha)}}{\ell(\mu)!}\,\LL  P^{(\alpha)}_\lambda| p_\mu\RR^{(\alpha)}.
\end{equation}
The Macdonald automorphism \eqref{defHomo} is then exploited for getting
\begin{equation}
 \left\langle \frac{P^{(\alpha)}_\lambda(x)}{P^{(\alpha)}_\lambda(1^N)}\right\rangle_{x\in\GbE_N}=(-1)^{\ell(\mu)}{2}^{\ell(\mu)}\frac{h^\lambda_{(\alpha)}}{\ell(\mu)!}\,\LL  P^{(\alpha)}_\lambda| \omega_{1/\alpha}p_\mu\RR^{(\alpha)}.
\end{equation}
Moreover, the use of Eq.\ \eqref{EqHomo} yields
\begin{multline}
 \left\langle \frac{P^{(\alpha)}_\lambda(x)}{P^{(\alpha)}_\lambda(1^N)}\right\rangle_{x\in\GbE_N}=(-1)^{\ell(\mu)}{2}^{\ell(\mu)}\frac{h^\lambda_{(\alpha)}}{\ell(\mu)!}\,\LL  P^{(\alpha)}_\lambda| \omega_{1}p_\mu\RR^{(1)}\\=(-1)^{\ell(\mu)}{2}^{\ell(\mu)}\frac{h^\lambda_{(\alpha)}}{\ell(\mu)!}\LL \omega_{\alpha} P^{(\alpha)}_\lambda| p_\mu\RR^{(1/\alpha)}.
\end{multline}
 But according to the duality Eq.\ \eqref{eqDualityJack} of the Jack polynomials, the last equation can be rewritten as
\begin{equation}
 \left\langle \frac{P^{(\alpha)}_\lambda(x)}{P^{(\alpha)}_\lambda(1^N)}\right\rangle_{x\in\GbE_N}=(-1)^{\ell(\mu)}{2}^{\ell(\mu)}\frac{h^\lambda_{(\alpha)}}{\ell(\mu)!} \frac{h^{\lambda'}_{(\alpha')}}{h_{\lambda'}^{(\alpha')}}\,\LL P^{(\alpha')}_{\lambda'}| p_\mu\RR^{(\alpha')},
\end{equation}
where  $\alpha'=1/\alpha$. Note that the last displayed equation is non-zero only if $N'\geq \ell(\lambda')$, where $N'$ denotes the number of variables in $P^{(\alpha')}_{\lambda'}$.  Besides, from the definition \eqref{eqHook} of the lower and upper hook-lengths, one easily obtains
\begin{equation}
 h^\lambda_{(\alpha)}=\alpha^{|\lambda|}h_{\lambda'}^{(\alpha')}=\alpha^{2\ell(\mu)}h_{\lambda'}^{(\alpha')}
\end{equation}
Hence,
\begin{equation}
 \left\langle \frac{P^{(\alpha)}_\lambda(x)}{P^{(\alpha)}_\lambda(1^N)}\right\rangle_{x\in\GbE_N}=(-1)^{\ell(\mu)}{2}^{\ell(\mu)} \alpha^{2\ell(\mu)}\frac{ h^{\lambda'}_{(\alpha')}}{\ell(\mu)!}\,\LL P^{(\alpha')}_{\lambda'}| p_\mu\RR^{(\alpha')}.
\end{equation}
Finally, Lemma 1 or Eq.\ \eqref{eqProofTheo1} is used once again to obtain
\begin{equation}
 \left\langle \frac{P^{(\alpha)}_\lambda(x)}{P^{(\alpha)}_\lambda(1^N)}\right\rangle_{x\in\GbE_N}={(-\alpha)^{\ell(\mu)}} \left\langle\frac{P^{(\alpha')}_{\lambda'}(y)}{P^{(\alpha')}_{\lambda'}(1^{N'})}\right\rangle_{y\in\GBE_{N'}}.
\end{equation}
which is the desired result.
\end{proof}

The simplest example of such a duality is certainly the average of moments of the characteristic polynomial (cf. \cite{Normand}).  Indeed, from
\begin{equation}
 (\det \X)^n = (x_1\ldots x_N)^n = P^{(2/\beta)}_{(n^N)}(x_1,\ldots,x_N)
\end{equation}
and the substitution of the the latter formula into Eq.\ \eqref{eqTheo1}, one concludes that
\begin{equation}
 \left\langle (\det \X)^n\right\rangle_{\X\in\GbE_N}=(-2/\beta)^{nN/2}\left\langle
 (\det \X)^N\right\rangle_{\X\in\GBE_{n}},
\end{equation}
for $\beta=4/\beta'=1,2,4$.  More generally,  for all $\beta>0$,
\begin{equation}
 \left\langle (x_1\cdots x_N)^n\right\rangle_{x\in\GbE_N}=(-2/\beta)^{nN/2}\left\langle
 (x_1\cdots x_n)^N\right\rangle_{x\in\GBE_{n}}.
\end{equation}

Another simple manifestation of Proposition \ref{theo1} is related to the
Hermite polynomials.  It is well know that an orthogonal polynomial of degree $N$ has a $N\times N$ matrix integral representation.  However, a classical orthogonal polynomial can also be realised as a  single integral.  For instance, the Hermite polynomial of degree $N$ can be written either as
\begin{equation}
 H_N(t)=2^N \Big\langle \prod_{i=1}^N(t-x_i)\Big\rangle_{x\in \mathrm{G2E}}= 2^N\Big\langle \det(t\I-\X)\Big\rangle_{\X\in \mathrm{G2E}}
\end{equation}
or as
\begin{equation}
 H_N(t)=\frac{2^N}{\sqrt{\pi}}\int_{-\infty}^\infty
 dx\,e^{-x^2}(t\pm\mathrm{i}x)^N.
\end{equation}
 Proposition \ref{theo1} allows to go directly from one representation to the other without referring to the orthogonality, determinantal representation, or differential equation.  On the one hand,
\begin{equation}
 \prod_{i=1}^N(t-x_i)=\sum_{n=0}^{N}t^{N-n}(-1)^{n}e_n(x_1,\ldots,x_N)=\sum_{n=0}^{N}t^{N-n}(-1)^{n}P^{(\alpha)}_{(1^n)}(x_1,\ldots,x_N),
\end{equation}
where $\alpha=2/\beta$, and
\begin{equation}
 P^{(\alpha)}_{(1^n)}(1^N)=\binom{N}{n},
\end{equation}
so that
\begin{equation}
 \Big\langle \prod_{i=1}^N(t-x_i)\Big\rangle_{x\in \GbE_N}=\sum_{n=0}^{N}t^{N-n}(-1)^{n}\binom{N}{n}\left\langle \frac{P^{(\alpha)}_{(1^n)}(x)}{ P^{(\alpha)}_{(1^n)}(1^N)}\right\rangle_{\GbE_N}.
\end{equation}
On the other hand, Proposition 1 and $(1^n)'=(n)$ immediately imply that the last equation is equal to
\begin{equation}
 \sum_{n=0}^{N}t^{N-n}(-1)^{n}\binom{N}{n}\left\langle \frac{P^{(1/\alpha)}_{(n)}(\mathrm{i}\sqrt{\alpha}x)}{ P^{(1/\alpha)}_{(n)}(1)}\right\rangle_{\GBE_1}= \sum_{n=0}^{N}t^{N-n}(-1)^{n}\binom{N}{n}\frac{1}{\sqrt{\pi}}\int_\mathbb{R}dxe^{-x^2}\left(\mathrm{i}\sqrt{\alpha}x\right)^n
\end{equation}
Consequently, for all $\beta=2/\alpha>0$,
\begin{equation}
 \Big\langle \prod_{i=1}^N(t-x_i)\Big\rangle_{x\in \GbE_N}=\frac{1}{\sqrt{\pi}}\int_{-\infty}^\infty
 dx\,e^{-x^2}(t-\mathrm{i}\sqrt{\alpha}x)^N=\left(\frac{\sqrt{\alpha}}{2}\right)^NH_N(t/\sqrt{\alpha})
\end{equation}

\subsection{Chiral ensembles}

Lemma \ref{lemma1} and Proposition \ref{theo1} can be easily adapted to the chiral case.
\begin{lemma}
Let $\mu$ stand for the partition $(1^{|\lambda|})$, i.e., $\ell(\mu)=|\lambda|$.   Then, for all partition $\lambda$, $\beta>0$,  and $\gamma>-1$,
\begin{equation}\label{EqProof2}
\left\langle {P^{(2/\beta)}_\lambda(x)}\right\rangle_{x\in\LbE^\gamma_N}=[\gamma+q]^{(2/\beta)}_\lambda\left(\frac{\beta}{2}\right)^{|\lambda|}\frac{b_\lambda^{(2/\beta,N)}}{|\lambda|!}\LLb
 P^{(2/\beta)}_\Lambda\,\Big|\,p_{\mu}\RRb^{(2/\beta)},
\end{equation}
where $q=1+\beta(N-1)/2$.  Equivalently,
\begin{equation}
\left\langle {P^{(2/\beta)}_\lambda(x)}\right\rangle_{x\in\LbE^\gamma_N}=[\gamma+q]^{(2/\beta)}_\lambda{b_\lambda^{(2/\beta,N)}}\underset{p_\mu}{\mathrm{coeff}}
 P^{(2/\beta)}_\Lambda.
\end{equation}
\end{lemma}
\begin{proof}  The orthogonality \eqref{eqOrthoL} and Eq.\ \eqref{eqLaguerre} give \cite[Proposition 4.11]{Baker}
\begin{equation}\label{eqProof2}
 \left\langle{\,_0{\mathcal{F}}_1}^{(\alpha)}(\gamma+q;x;y){\,_0{\mathcal{F}}_1}^{(\alpha)}(\gamma+q;x;z) \right\rangle_{x\in\LbE}=e^{p_1(y)+p_1(z)}{\,_0{\mathcal{F}}_1}^{(\alpha)}(\gamma+q;y;z)
\end{equation}
for $\alpha=2/\beta$.  Then, one sets $z=0$ and proceeds as in Lemma \ref{lemma1}.
\end{proof}

Employing the same method as that exposed in the proof of Proposition \ref{theo1}, one shows that the last lemma and the duality \eqref{eqDualityJack} for the Jack polynomials imply the following proposition.

\begin{proposition} \label{theo2}
Set ${\beta'}=4/\beta>0$, $q=1+(N-1)\beta/2$, and $q'=1+(N'-1)2/\beta$. Moreover, assume that $N\geq\ell(\lambda)$, $N'\geq \ell(\lambda')$, $\gamma>-1$, and $\gamma'>-1$.
  Then,
\begin{equation}\label{eqProofTheo2}
\left\langle \frac{P^{(2/\beta)}_\lambda(x)}{P^{(2/\beta)}_\lambda(1^{N})}\right\rangle_{x\in\LbE^\gamma_N}=\,\frac{[\gamma+q]^{(2/\beta)}_\lambda}{[\gamma'+q']^{(2/\beta')}_{\lambda'}}\,
\left\langle
 \frac{P^{(2/\beta')}_{\lambda'}(x)}{P^{(2/\beta')}_{\lambda'}(1^{N'})}\right\rangle_{x\in\LBE^{\gamma'}_{N'}}.
\end{equation}
\end{proposition}

As an example, choose $\lambda=(n^N)$ and $\gamma'=2(\gamma+1)/\beta-1$.  Direct manipulations then give
\begin{equation}
 \frac{{[\gamma+q]^{(2/\beta)}_\lambda}}{{[\gamma'+q']^{(2/\beta')}_{\lambda'}}}=\left(\frac{\beta}{2}\right)^{nN}.
\end{equation}
This implies that moments of the determinant also satisfy a simple duality in the Chiral $\beta$-Ensemble:
\begin{equation}\
\left\langle (x_1\cdots x_N)^n\right\rangle_{x\in\LbE^\gamma_N}=\left(\frac{\beta}{2}\right)^{nN}\left\langle (x_1\cdots x_n)^N\right\rangle_{x\in\LBE^{\gamma'}_N}.
\end{equation}

It is worth mentioning that the average in the Gaussian ensembles are limit cases of similar averages in the Chiral ensembles.  This can be understood as follows. Set $y=\gamma+\sqrt{2\gamma}x$ with $\gamma>0$.  Trivial manipulations give
\begin{equation}
\gamma^{-1}e^\gamma y^\gamma e^{-y}=e^{-x^2}+\mathcal{O}(\gamma^{-1/2})
\end{equation}
uniformly when $\gamma$ goes to $\infty$.  Then, simple changes of variable allows one to conclude that
\begin{equation}\lim_{\gamma\rightarrow \infty} \left\langle F(-\sqrt{\gamma/2}+y/\sqrt{2\gamma})\right\rangle_{y\in\LbE^\gamma_N}=\left\langle F(x)\right\rangle_{x\in\GbE_N}
\end{equation}
for all multivariate polynomials $F$, positive integers $N$ and $\beta>0$.  For instance, by setting $\gamma'=\gamma$ and taking the above limit in both sides of Proposition \ref{theo2}, one readily establishes Proposition~\ref{theo1}.

\section{Source -- external field dualities}

In this section, averages of (inverse) characteristic polynomials in $\beta$-Ensembles with external fields are studied.  Central to the analysis is the use of Dunkl transforms (see for instance \cite{Rosler}).

\subsection{Gaussian ensembles}

The following formula is a generalisation of the Fourier transformation \cite{Baker,Rosler}:
\begin{equation}\label{eqFourier}
 e^{-p_2(y)} \left\langle{\,_0{\mathcal{F}}_0}^{(2/\beta)}(2x;y)F(x)\right\rangle_{x\in\GbE}=e^{\frac{1}{4}\Delta^{(2/\beta)}_y}F(y)
\end{equation}
where
\begin{equation}
 \Delta^{(\alpha)}_x=\sum_{i=1}^N\left(\frac{\partial}{\partial {x_i}}+\frac{2}{\alpha}\sum_{j\neq i}\frac{1}{x_i-x_j}\right)\frac{\partial}{\partial {x_i}}.
\end{equation}
 The equation holds true for any analytic function $F$.  It can be seen as a consequence of Eq.\ \eqref{EqProof1} and the Lassalle formula
\begin{equation}\label{eqLassalle}
 \mathcal{H}_\lambda^{(\alpha)}(x)=e^{-\frac{1}{4}\Delta_x^{(\alpha)}}P^{(\alpha)}_\lambda(x).
\end{equation}

Now imagine that there exists a function $G(s;f)$ of two sets of variables, $s=(s_1,\ldots,s_n)$ and $f=(f_1,\ldots,f_N)$, satisfying for some $\alpha'$
\begin{equation}
 \Delta_f^{(\alpha)}G(s;f)=\Delta_s^{(\alpha')}G(s;f).
\end{equation}
Due to the commutativity of $\Delta_f^{(\alpha)}$ and $\Delta_s^{(\alpha')}$, this means that
\begin{equation}
 e^{\frac{1}{4}\Delta_f^{(\alpha)}}G(s;f)=e^{\frac{1}{4}\Delta_s^{(\alpha')}}G(s;f).
\end{equation}
Therefore, as a direct consequence of the latter formula and  Eq.\ \eqref{eqFourier},
\begin{multline}
  e^{-p_2(f)} \left\langle{\,_0{\mathcal{F}}_0}^{(2/\beta)}(2x;f)G(s;x)\right\rangle_{x\in\GbE}= e^{-p_2(s)} \left\langle{\,_0{\mathcal{F}}_0}^{(2/\beta')}(2y;f)G(y;f)\right\rangle_{y\in\GBE}
\end{multline}
where it is understood that $\beta'=2/\alpha'$.

\begin{proposition} \label{prop3} Set $\beta'=4/\beta$.
  Then, for all positive integers $n$ and $N$, and all $\alpha>0$,
\begin{multline}\label{eq1prop3}
e^{-p_2(f)} \left\langle \prod_{j=1}^n\prod_{k=1}^N\left(s_j\pm \mathrm{i}\sqrt{\frac{2}{\beta}}x_k\right) \,_0{\mathcal{F}_0}^{(2/\beta)}(x,2f) \right\rangle_{x\in\GbE_N} = \\
e^{-p_2(s)} \left\langle \prod_{j=1}^n\prod_{k=1}^N \left(y_j\pm \mathrm{i}\sqrt{\frac{2}{\beta}}f_k\right) \,_0{\mathcal{F}_0}^{(2/\beta')}(y,2s)\right\rangle_{y\in\GBE_n}.
\end{multline}
If in addition the variables $s$ and $f$ are not real, then
\begin{multline}\label{eq2prop3}
e^{-p_2(f)} \left\langle \prod_{j=1}^n\prod_{k=1}^N\left(s_j\pm x_k\right)^{-\beta/2} \,_0{\mathcal{F}_0}^{(2/\beta)}(x,2f) \right\rangle_{x\in\GbE_N} = \\
e^{-p_2(s)} \left\langle \prod_{j=1}^n\prod_{k=1}^N \left(y_j\pm f_k\right)^{-\beta/2} \,_0{\mathcal{F}_0}^{(2/\beta)}(y,2s)\right\rangle_{y\in\GbE_n}.
\end{multline}
Remark that in the last equation, $\beta$ is not affected by the duality transformation.
\end{proposition}
\begin{proof}
Recall that  $\alpha=2/\beta$ and let
\begin{equation}
 \prod(s;af)^b=\prod_{j=1}^n\prod_{k=1}^N\left(s_j-af_k\right)^b.
\end{equation}
Following the above discussion on the Dunkl transform, it is sufficient prove that
\begin{equation}
 \Delta^{(\alpha)}_f \prod(s;af)^b= \Delta^{(\alpha')}_s \prod(s;af)^b
\end{equation}
for appropriate values of  $a$, $b$, and $\alpha'$.  On the one hand, by using
\begin{equation}
 \frac{\partial}{\partial f_k}  \prod(s;af)^b=-ab\sum_i\frac{1}{s_i-af_k} \prod(s;af)^b
\end{equation}
and
\begin{equation}
 \sum_{i}\sum_{k\neq l}\frac{1}{s_i-af_k}\frac{1}{f_k-f_l}=\frac{a}{2}\sum_i\sum_{k\neq l}\frac{1}{s_i-af_k}\frac{1}{s_i-af_l},
\end{equation}
it is simple to show that
\begin{multline}\label{eqDeltaf}
  \Delta^{(\alpha)}_f \prod(s;af)^b=\sum_{i,k}\frac{1}{s_i-af_k}\left( a^2b(b-1)\frac{1}{s_i-af_k}\phantom{\frac{a^2b}{\alpha}} \right. \\
 \left.+a^2b^2\sum_{j\neq i}\frac{1}{s_j-af_k}
-\frac{a^2b}{\alpha}\sum_{l\neq k}\frac{1}{s_i-af_l}\right)\prod(s;af)^b
\end{multline}
On the other hand, exploiting
\begin{equation}
 \frac{\partial}{\partial s_k}  \prod(s;af)^b= b\sum_k\frac{1}{s_i-af_k} \prod(s;af)^b
\end{equation}
and
\begin{equation}
 \sum_{i\neq j}\sum_{k}\frac{1}{s_i-af_k}\frac{1}{s_i-s_j}=-\frac{1}{2}\sum_{i\neq j}\sum_{k}\frac{1}{s_i-af_k}\frac{1}{s_j-af_k},
\end{equation}
one finds
\begin{multline}\label{eqDeltas}
  \Delta^{(\alpha')}_s \prod(s;af)^b=\sum_{i,k}\frac{1}{s_i-af_k}\left( b(b-1)\frac{1}{s_i-af_k}\phantom{\frac{a^2b}{\alpha}} \right. \\
 \left.-\frac{b}{\alpha'}\sum_{j\neq i}\frac{1}{s_j-af_k}
+b^2\sum_{l\neq k}\frac{1}{s_i-af_l}\right)\prod(s;af)^b
\end{multline}
Note that $s$ and $f$ are kept generic.  Thus, imposing the equality of Eqs.\ \eqref{eqDeltaf} and \eqref{eqDeltas} requires
\begin{equation}
 a^2(b-1)=(b-1),\quad \alpha' a^2b=-1, \quad a^2=-b\alpha.
\end{equation}
The only solutions to the latter system of equations are either
\begin{equation}
 a=\pm \mathrm{i}\sqrt{\alpha},\quad b=1,\quad \alpha'=1/\alpha.
\end{equation}
or
\begin{equation}
 a=\pm1,\quad b=-1/\alpha,\quad \alpha'=\alpha.
\end{equation}
This completes the proof of the proposition.
\end{proof}

Consider for example the expectation value of a product of characteristic polynomials without external field, i.e., for  $f=0$.  According to Proposition \ref{prop3},
\begin{equation}
 \left\langle \prod_{j=1}^n\prod_{k=1}^N\left(s_j\pm \sqrt{\frac{2}{\beta}}x_k\right)  \right\rangle_{x\in\GbE_N} =
e^{p_2(s)} \left\langle \prod_{j=1}^n (\mathrm{i}y_j)^N \,_0{\mathcal{F}_0}^{(2/\beta')}(2y,-\mathrm{i}s)\right\rangle_{y\in\GBE_n}.
\end{equation}
The right-hand side can  be written as
\begin{equation}
 e^{p_2(s)} \left\langle P^{(\beta/2)}_{(N^n)}(\mathrm{i}y)\,_0{\mathcal{F}_0}^{(2/\beta')}(2y,\mathrm{i}s)\right\rangle_{y\in\GBE_n}.
\end{equation}
Due to the Dunkl transform \eqref{eqFourier}, the latter equation is also equal to $\exp\left(-\frac{1}{4}\Delta^{(\beta/2)}_s\right)P^{(\beta/2)}_{(N^n)}(s)$.  Returning to the Lassalle formula \eqref{eqLassalle}, it becomes clear that  averages of products of characteristic polynomials are multivariate Hermite polynomials:
\begin{equation}\label{eqHprod}
 \mathcal{H}^{(\beta/2)}_{(N^n)}(s)= \left\langle \prod_{j=1}^n\prod_{k=1}^N\left(s_j\pm\sqrt{\frac{2}{\beta}}x_k\right)  \right\rangle_{x\in\GbE_N}.
\end{equation}
 This equivalence was first pointed out  in \cite{Baker} by considering a limit of the multivariate Jacobi polynomial.  Another proof consists in showing that the expectation value satisfies a Calogero-like eigenvalue problem.

Propositions \ref{prop1} and \ref{prop2} given in the Introduction are in fact special cases of Proposition \ref{prop3}. The connection between the averages involving multivariate hypergeometric functions and the matrix integrals can be understood as follows.  Let $\beta=1,2,$ or $4$,  and let $(\U^\dagger d\U)$ stand for the  normalised Haar measure for unitary matrices with real, complex or quaternion real elements respectively.  It is a standard result \cite{ForresterBook,Mehta} that
\begin{equation}
(d\X)\,F(\X)=\frac{1}{C} \prod_{i=1}^N dx\prod_{1\leq i<j\leq N}|x_i-x_j|^\beta (\U^\dagger d\U)F(\U x\U^\dagger).
\end{equation}
for any Hermitian $\beta$-matrix $\X=\U x \U^\dagger$, where $x=\mathrm{diag} (x_1,\ldots,x_N)$, and some constant $C$.  When $F$ is not invariant, i.e., when $F(\U x\U^\dagger)\neq F(x)$, the calculation of the expectation values containing $F$ requires the use of  the theory of zonal polynomials (see Chapter VII in Macdonald's classical book \cite{Mac}).  One can show for instance that, if $\X$ and $\Y$ are $N\times N$ Hermitian $\beta$-matrices,
\begin{equation}
 \int (\U^\dagger d\U) P^{(2/\beta)}_\lambda (\X \U \Y \U^\dagger)=\frac{P^{(2/\beta)}_\lambda(x) P^{(2/\beta)}_\lambda(y)}{P^{(2/\beta)}_\lambda(1^N)}.
\end{equation}
In the last equation,  $P^{(2/\beta)}_\lambda (\X \U \Y \U^\dagger)=P^{(2/\beta)}_\lambda (z_1,\ldots,z_N)$, where  $(z_1,\ldots,z_N)$ denotes the eigenvalues of $\X \U \Y \U^\dagger$.  Now, by exploiting formula \eqref{eqHyperspecial}, one readily gets
\begin{equation}\label{angularint}
 \int (\U^\dagger d\U) e^{\tr (\X \U \Y \U^\dagger)}=\,_0{\mathcal{F}_0}^{(2/\beta)}(x;y),
\end{equation}
and consequently, for any invariant function $g(\X)=g(x)$ and $\beta=1, 2$ or $4$,
\begin{multline}
 \left\langle \,_0{\mathcal{F}_0}^{(2/\beta)}(x;f)  g(x)\right\rangle_{x\in\GbE_N}=\\
{\int (d\X)e^{-\tr \X^2+\tr \X\F}g(\X)}\Big/{\int (d\X)e^{-\tr \X^2}}= \left\langle e^{\tr \X\F} g(\X)\right\rangle_{\X\in\GbE_N}.
\end{multline}
Finally, the comparison of the latter formula with Eqs.\ \eqref{eq1prop3} and \eqref{eq2prop3} respectively establishes Propositions \ref{prop1} and \ref{prop2}.

\subsection{Chiral ensembles}

There exists a Lassalle formula for the multivariate Laguerre polynomials.  For $\alpha=2/\beta>0$ and $\gamma>-1$, it reads \cite{Baker}
\begin{equation}\label{eqLassalle2}
 \mathcal{L}_\lambda^{(\alpha,\gamma)}(x)=e^{-\Delta_x^{(\alpha,\gamma)}}P^{(\alpha)}_\lambda(x),
\end{equation}
where
\begin{equation}\label{Delta}
 \Delta^{(\alpha,\gamma)}_x=\sum_{i=1}^N\left(x_i\frac{\partial}{\partial {x_i}}+\frac{2}{\alpha}\sum_{j\neq i}\frac{x_i}{x_i-x_j}+\gamma+1\right)\frac{\partial}{\partial {x_i}}.
\end{equation}
From formula \eqref{eqLassalle2} and Eq.\ \eqref{eqProof2}, one can prove the following  generalisation
 of the Hankel transform which is due to Dunkl \cite{Baker, Rosler} :
\begin{equation}\label{eqHankel}
 e^{p_1(y)} \left\langle {\,_0{\mathcal{F}}_1}^{(2/\beta)}(\gamma+q;x;-y)F(-x) \right\rangle_{x\in\LbE_N^\gamma}=e^{-\Delta^{(2/\beta,\gamma)}_y}F(y)
\end{equation}
with  and $q=1+\beta(N-1)/2$.   The strategy for getting dualities in the Chiral $\beta$-Ensemble is the same as in the Gaussian $\beta$-Ensemble; that is, find a function $G(s;f)$ such that
\begin{equation}
 \Delta_f^{(\alpha,\gamma)}G(s;f)=\Delta_s^{(\alpha',\gamma')}G(s;f),
\end{equation}
and exploit the Dunkl transform \eqref{eqHankel} in order to conclude that
\begin{multline}
  e^{p_1(f)} \left\langle{\,_0{\mathcal{F}}_1}^{(2/\beta)}(\gamma+q;x;-f)G(s;-x)\right\rangle_{x\in\LbE_N^\gamma}= \\
e^{p_1(s)} \left\langle{\,_0{\mathcal{F}}_1}^{(2/\beta')}(\gamma'+q';y;-s)G(-y;f)\right\rangle_{y\in\LBE^{\gamma'}}
\end{multline}
for some $\beta'$, $\gamma'$, and $q'=1+\beta'(n-1)/2$.

\begin{proposition} \label{prop4} Set $\beta'=4/\beta$, $\gamma'=2(\gamma+1)/\beta-1$.
  Then, for all positive integers $n$ and $N$,  all $\alpha>0$ and $\gamma>-1$,
\begin{multline}\label{eq1prop4}
e^{p_1(f)} \left\langle \prod_{j=1}^n\prod_{k=1}^N\left(s_j-{\frac{2}{\beta}}x_k\right) \,_0{\mathcal{F}_1}^{(2/\beta)}(\gamma+q;x,-f) \right\rangle_{x\in\LbE_N^\gamma} = \\
(-1)^{nN}e^{p_1(s)} \left\langle \prod_{j=1}^n\prod_{k=1}^N \left(y_j-{\frac{2}{\beta}}f_k\right) \,_0{\mathcal{F}_1}^{(2/\beta')}(\gamma'+q';y,-s)\right\rangle_{y\in\LBE_n^{\gamma'}}.
\end{multline}
Suppose moreover that $s$ and $f$ have non-zero imaginary parts, then
\begin{multline}\label{eq2prop4}
e^{p_1(f)} \left\langle \prod_{j=1}^n\prod_{k=1}^N\left(s_j+ x_k\right)^{-\beta/2} \,_0{\mathcal{F}_1}^{(2/\beta)}(\gamma+q;x,-f) \right\rangle_{x\in\LbE_N^\gamma} = \\(-1)^{\beta nN/2}
e^{p_1(s)} \left\langle \prod_{j=1}^n\prod_{k=1}^N \left(y_j+ f_k\right)^{-\beta/2} \,_0{\mathcal{F}_1}^{(2/\beta)}(\gamma'+q';y,-s)\right\rangle_{y\in\GbE_n^{\gamma'}}.
\end{multline}
The latter equation is valid only if $\beta>2\gamma$, which ensures  $\gamma'>-1$.
\end{proposition}
\begin{proof}
Define
\begin{equation}
 \prod(s;af)^b=\prod_{j=1}^n\prod_{k=1}^N\left(s_j-af_k\right)^b.
\end{equation}
Direct calculations lead to the conclusion that
\begin{equation}
 \Delta^{(\alpha,\gamma)}_f \prod(s;af)^b= \Delta^{(\alpha',\gamma')}_s \prod(s;af)^b
\end{equation}
 for all $n$ and $N$, if and only if,
\begin{equation}
 ab\alpha'=-1,\quad a(b-1)=(b-1),\quad a(\gamma+1)=-(\gamma'+b),\quad a=-b\alpha.
\end{equation}
The proposition follows from the fact that the unique solutions to these equations are
\begin{equation}
 a=-\alpha,\quad b=1,\quad \alpha'=1/\alpha, \quad \gamma'=\alpha(\gamma+1)-1
\end{equation}
and
\begin{equation}
  a=1,\quad b=-1/\alpha,\quad \alpha'=\alpha, \quad \gamma'=1/\alpha -\gamma-1.
\end{equation}

\end{proof}

The comparison of Eqs.~\eqref{eqLassalle2}, \eqref{eqHankel}, and \eqref{eq1prop4}  leads to a simple generalisation of Eq.\ \eqref{eqHprod}, that is, a random matrix representation of the multivariate Laguerre polynomial:
\begin{equation}
\mathcal{L}^{(2/\beta',\gamma')}_{(N^n)}(s)= \left\langle \prod_{j=1}^n\prod_{k=1}^N\left(s_j-{\frac{2}{\beta}}x_k\right)  \right\rangle_{x\in\LbE^\gamma_N},
\end{equation}
where $\beta'=4/\beta$ and $\gamma'=2(\gamma+1)\beta-1$.

\section{Further applications}

\subsection{Multiple polynomials}

 Ensembles of random Hermitian matrices   with external field naturally lead to multiple polynomials  when $\beta=2$.  These polynomials satisfy orthogonality conditions for more than one scalar products; they can be interpreted as  multi-parameter deformations of the usual orthogonal polynomials in one variable (see the review by van~Assche in \cite{vAssche}).

It is known that the multiple Hermite polynomials of type  II can be defined as follows: \footnote{The factor $\exp(-\tr \F^2/4)$ comes from the absence of $\F$ in the definition \eqref{Mgbe} of p.d.f.\ for GUE.  See also Eq.\  \eqref{eqGauss}.}
\begin{equation} \label{mHermiteND}
 \mathcal{H}_\mathbf{n}(z)=e^{-\frac{1}{4}\tr \F^2}\left\langle \det(z\I-\X)e^{\tr \X\F}\right\rangle_{\X\in \mathrm{GUE}_N}
\end{equation}
Here the polynomial is chosen monic of degree $N$ while  $\mathbf{n}=(n_1,\ldots,n_D)$ is the multiplicity vector for the eigenvalues of $\F$.  In other words, $\F$ has $n_i$ eigenvalues equal to $g_i$ say, and $\sum_in_i=N$. One can write for instance
\begin{equation}
 f=g^\mathbf{n}=(g_1^{n_1},\ldots,g_D^{n_D}).
\end{equation}
 As a consequence of Proposition 1, it is clear that the multiple Hermite polynomial also has  the following realisation:
\begin{equation}\label{mHermite1D}
 \mathcal{H}_\mathbf{n}(z)=\frac{(-\mathrm{i})^N}{\sqrt{\pi}}\int_{\mathbb{R}} dy\,e^{-(y-\mathrm{i}z)^2}\prod_{j=1}^D(y-\mathrm{i}g_j/2)^{n_i}.
\end{equation}
This formula has been first obtained by Bleher and Kuijlaars in \cite{Bleher} from the orthogonality relations satisfied by $P_{\mathbf{n}}$.

There is another interesting application of Proposition 1.  Indeed, the comparison of Eqs.\ \eqref{DualMat1} and \eqref{mHermite1D} indicates that the multiple Hermite polynomials can be represented as $\beta$-matrix integrals
\begin{equation}
  \mathcal{H}_\mathbf{n}(z)=e^{-\frac{1}{4}\frac{\beta}{2}\tr \F^2}\left\langle \det\left(z\I-\sqrt{\frac{2}{\beta}}\X\right)e^{\sqrt{\frac{\beta}{2}}\tr \X\F}\right\rangle_{\X\in \GbE_N}
\end{equation}
for $\beta=1,2,4$, or as
\begin{equation}
 \mathcal{H}_\mathbf{n}(z)=e^{-\frac{1}{4}\frac{\beta}{2}p_2(f)}\left\langle \prod_{j=1}^N\left(z-\sqrt{\frac{2}{\beta}}x_j\right)\,_0{\mathcal{F}_0}^{(2/\beta)}\left(x;\sqrt{\frac{\beta}{2}}f\right) \right\rangle_{x\in \GbE_N}
\end{equation}
for all $\beta>0$.

The multiple Hermite function of type I can also be written as a single integral \cite{Bleher}.
\begin{equation}
 Q_{\mathbf{n}}(z)=\lim_{\epsilon\rightarrow 0^+}\frac{1}{\pi}\Im  \int_\mathbb{R} dy\,e^{-(y-z+\mathrm{i}\epsilon)}\prod_{j=1}^D(y-f_j/2)^{n_j} .
\end{equation}
Note that taking the imaginary part is equivalent to closing the contour around the poles.  Now suppose that $\beta/2$ is a positive integer and that $n_i=\beta m_i/2$ for all $i=1,\ldots,D$.  In addition, construct a reduced sequence of variables  (only the multiplicities are changed)
\begin{equation}
 \tilde{f}=g^\mathbf{m}=(g_1^{m_1},\ldots,g_D^{m_D}).
\end{equation}
 and set $M=\sum_j m_j=2N/\beta$.  Then Proposition 2 furnishes new integral representations for the multiple Hermite functions of type II:
\begin{equation}
  Q_{\mathbf{n}}(z)=e^{-\frac{1}{4}p_2(\tilde{f})} \lim_{\epsilon\rightarrow 0^+}\frac{1}{\pi}\Im
\left\langle \prod_{j=1}^M(z-\mathrm{i}\epsilon-x_j)^{-\beta/2}\,_0{\mathcal{F}_0}^{(2/\beta)}(x;\tilde{f})\right\rangle_{x\in\GbE_M}.
\end{equation}

Similar representations can be deduced for the multiple Laguerre polynomials of type I and II by exploiting the results given in Section 4.2.

\subsection{Formal one-matrix models}

One point about matrix models encountered in theoretical physics and combinatorics deserves to be clarified.

Matrix integrals are often used for solving combinatorial problems related to 2D Quantum Gravity, such as counting graphs drawn on surfaces of a given genus (for a simple but precise introduction to the subject, see \cite{Zvonkin}; more advanced topics and recent references can be found in \cite{DiFrancesco}).  Questions of convergence of the integrals are not an issue when  enumerating objects; only formal power series are considered.  One typically looks at  functions involving traces of a random matrix averaged over the Gaussian ensemble of Hermitian matrices (usually $\beta=2$).   Averages are obtained from the partition function of the one-matrix model:
\begin{equation}\label{FormalPart}
 Z_{N,\beta}(t_1,t_2\ldots)=\int (d\X) \exp\left( -\sum_{k\geq 1} \frac{t_k}{k}\tr \X^k\right).
\end{equation}

For $\beta=2$, Chekov and Makeenko \cite{Chekov} have shown, by exploiting Schwinger-Dyson equations, the equivalence of the one-matrix model to the Gaussian model with an external field.  In fact, this remains true for $\beta=1$ and $4$. It can be proved quickly.  First, set
\begin{equation}\label{eqt}
 t_1=-2u+v\tr\bS^{-1},\quad t_2=2+v^2\tr\bS^{-2},\quad\mbox{and}\quad t_k= v^k\tr \bS^{-k} \quad\mbox{for}\quad k\geq3,
\end{equation}
 where u is a formal parameter,  $v=\pm\mathrm{i}\sqrt{2/\beta}$, and $\bS$ is a $n \times n$ Hermitian or anti-Hermitian whose eigenvalues $s$ are non-zero (so, $t_k$ is essentially a $k$th power sum).  Then, the formal development of  $\sum_{k\geq 1} {t_k}\tr \X^k/k$ in terms of the eigenvalues of $\bS$ yields
\begin{equation}\label{eqFormalProd}
 \exp\left( -\sum_{k\geq 1} \frac{t_k}{k}\tr \X^k\right)=(\det \bS)^{-N}\prod_{j=1}^n\det(s_j\I-v\X)\exp\left( -\tr \X^2+2u\tr \X \right).
\end{equation}
Finally, from the substitution of the last equation in Eq.\ \eqref{FormalPart} and the comparison with Proposition 1, it holds that the  formal one-matrix model is equivalent to a Gaussian model with an external field.

\begin{proposition}\label{propFormal} Let $\X$ be a   $N\times N$ Hermitian $\beta$-matrix.  Moreover, suppose that $\beta'=4/\beta$ and that $\bS$ is a $n\times n$ Hermitian $4/\beta$-matrix satisfying Eq.\ \eqref{eqt}.  Then
\begin{equation}
Z_{N,\beta}(t_1, t_2,\ldots)=
z_{N,\beta} \frac{e^{-\tr \bS^2+Nu^2}}{\det
\bS^{N}}\left\langle\det\left(\Y-\mathrm{i}\sqrt{\frac{\beta}{2}}u\I\right)^Ne^{2\tr \Y\bS}\right\rangle_{\Y\in\mathrm{G}\beta'\mathrm{E}_n},
\end{equation}
where
\begin{equation}
 z_{N,\beta}=\int (d\X) e^{-\tr \X^2}= \pi^{N/2}(\pi/2)^{\beta N(N-1)/4}.
\end{equation}
\end{proposition}

The latter result remains true for all $\beta>0$ if one appropriately uses the generalised hypergeometric functions $\,_0\mathcal{F}_0(2y;s)$ in place of $e^{2\tr \Y\bS}$ (cf. Section 4.1).

\subsection{Matrices at the edge of the spectrum}

Consider a random matrix belonging to the $\GbE_N$ where  $\beta=1,2$ or $4$.  Suppose moreover that the size $N$ of the matrix goes to infinity. Then, a classical result \cite{ForresterEdge} says that that the eigenvalue correlation functions, when recentered and rescaled at the spectrum  edge (i.e., $x_i\sim \sqrt{2N}$), can be expressed in terms of determinants (or Pfaffians) involving the Airy function and its derivative.  Here it is proved that the average of products of characteristic polynomials can be expressed, at the edge of the spectrum, as a matrix Airy integral.
For $\beta=2$, this integral has been first introduced by Kontsevich in \cite{Kontsevich} when studying the asymptotic behaviour of the one-matrix model's partition function \eqref{FormalPart}.

Let  $\W$ and $\F$ be $N\times N$ Hermitian $\beta$-matrices.  The matrix integral of the Kontsevich type is defined as follows:
\begin{equation}\label{matrixAiry}
 \mathrm{Ai}^{(2/\beta)}(\F)=a^{(2/\beta)}_N\int (d\W) \exp \left( \frac{\mathrm{i}}{3}\tr \W^3+\mathrm{i}\tr \W\F \right),
\end{equation}
where $a^{(\alpha)}_N=({2\pi})^{-N- N(N-1)/\alpha}$.  Obviously, the standard Airy function is recovered for $N=1$. Recall that the latter function satisfies $\mathrm{Ai}''(x)=x\mathrm{Ai}(x)$. From the derivation of Eq.\ \eqref{matrixAiry} and the use of integration by parts,  a simple generalisation of the Airy differential equation is obtained:
\begin{equation}
 \Delta_\F\,  \mathrm{Ai} (\F)= \tr \F \,\mathrm{Ai} (\F).
\end{equation}
In the last equation, $\Delta_\F$ stands for the Laplacian of the matrix $\F$; that is, 
\begin{equation}
 \Delta_{\F}=\sum_{i=1}^N \left(\frac{\partial}{\partial {F^0_{ii}}}\right)^2+\frac{1}{2}\sum_{a=0}^{\beta-1}\sum_{1\leq i<j\leq N}\left(\frac{\partial}{\partial {F^a_{ij}}}\right)^2.
\end{equation}

In fact, the matrix integral \eqref{matrixAiry} is a special realisation of a more general multivariate Airy function (see   Eq.\ \eqref{angularint}),
\begin{equation}\label{generalAiry}
 \mathrm{Ai}^{(\alpha)}(f)=a^{(\alpha)}_N \int_{\mathbb{R}^N}e^{\frac{\mathrm{i}}{3}\sum_jw_j^3} {\,_0\mathcal{F}_0}^{(\alpha)}(w;\mathrm{i}f) \prod_{1\leq j<k\leq N} |w_j-w_k|^{2/\alpha} \, dw ,
\end{equation}
 which is valid for any $\alpha=2/\beta>0$.  When $f_1=\ldots=f_N=x$ say, the latter function is equivalent to that previously introduced in \cite{DF} as the limiting eigenvalue density at the soft edge for the Chiral and Gaussian $\beta$-Ensembles ($\beta$  even).  For general $f=(f_1,\ldots,f_N)$ and $\alpha>0$, one has
\begin{equation}
 \Delta^{(\alpha)}_f\, \mathrm{Ai}^{(\alpha)}(f)= p_1(f)\, \mathrm{Ai}^{(\alpha)}(f),
\end{equation}
where $ \Delta^{(\alpha)}_f$ is the differential operator defined in Eq.\ \eqref{Delta} and $p$ is a power sum (see Section 2.2).  This multivariate Airy differential equation follows from the use of the Calogero-like equation \cite{Baker}
\begin{equation}
 \Delta^{(\alpha)}_x \,_0\mathcal{F}_0(x;y)=p_2(y)\, \,_0\mathcal{F}_0(x;y)
\end{equation}
and simple manipulations in the integral of Eq.\ \eqref{generalAiry}.

According to Eq.\ \eqref{eqHprod}, a multivariate Hermite polynomial associated to  a rectangular  partition is equivalent to an average of products of characteristic polynomials.   By virtue of Proposition \ref{prop1}, the latter  quantity can be replaced by the expectation value of products of determinants in an ensemble of random matrices with an external field.  Explicitly,
\begin{multline}\label{eqHermiteMat}
\mathcal{H}^{(\beta/2)}_{(N^n)}(s)=\left\langle
\prod_{i=1}^n\det\left(s_j\I\pm\sqrt{\frac{2}{\beta}}\X\right)\right\rangle_{\X\in\GbE_N}\\
=e^{\tr \bS^2}\left\langle (\det \mathrm{i}\Y)^Ne^{-2\mathrm{i}\tr\Y\bS}\right\rangle_{\Y\in\GBE_n}
\end{multline}
if $\beta'=4/\beta=1,2$ or $4$.  The last integral representation suits perfectly for the large $N$ asymptotic analysis.  The following result generalises the well known asymptotic expansion of the Hermite polynomial $H_N(x)$ in terms of the Airy function.

\begin{proposition}\label{propEdge} Let $G_{N,\beta}(\bS)$ denote the first line of Eq.\ \eqref{eqHermiteMat}, where $\bS$ is a Hermitian $4/\beta$-matrix whose eigenvalues are given by $s=(s_1,\ldots,s_n)$. Then, as $N\rightarrow\infty$,
\begin{equation}\label{eqPropEdge}
C^{-1}e^{-N^{1/3}\,\tr\, \bS}\,G_{N,\beta}\left(\sqrt{2N}\I+\frac{1}{\sqrt{2N^{1/3}}}\bS\right)\sim \mathrm{Ai}^{(\beta/2)}(\bS)+\mathcal{O}\left(\frac{1}{N^{1/3}}\right).
\end{equation}where
\begin{equation}\label{eqC}
C=2^{n(n-1)/\beta}\left(\frac{Ne}{2}\right)^{nN/2}(2\pi N^{1/3})^{n/2+n(n-1)/\beta}.
\end{equation}
\end{proposition}

The correct proof of the proposition is given in Appendix B.  It is purely technical and relies on the multidimensional steepest descent method.  However, the asymptotic formula can be easily understood thanks to the following heuristic argument.  First, by rescaling the matrix $\Y$ in Eq.\ \eqref{eqHermiteMat}, one finds
\begin{equation}\label{eqGMatrix}
 G_{N,\beta}\left(\sqrt{2N}\I+\frac{1}{\sqrt{2N^{1/3}}}\bS\right)=D\int (d\Y) \exp\left( Nf(\Y)+N^{1/3}g(\Y,\bS)\right)
\end{equation}
where
\begin{equation}
 f(\Y)=\tr \left(-2\Y^2-4\mathrm{i}\Y+\ln \Y\right),\quad g(\Y,\bS)=-2\tr \left(\Y\bS+i\bS\right)
\end{equation}
and 
\begin{equation}
 D=\frac{(-2N)^{nN/2}(2N)^{n/2+n(n-1)/\beta}}{\pi^{n/2}(\pi/2)^{n(n-1)/\beta}}e^{\frac{1}{2N^{1/3}}\tr \bS^2}
\end{equation}
Second, the following change of variables is  made in order to eliminate the quadratic terms in $f(\Y)$:
\begin{equation}\label{changeW}
 \W=2N^{1/3}\left(\Y+\frac{\mathrm{i}}{2}\I\right). 
\end{equation}
Third, the function $f(\Y)$ is formally expanded in powers of $N^{-1/3}$. This yields 
\begin{multline}\label{EqHeuristic}
 C^{-1} e^{-N^{1/3}\tr \bS-\frac{1}{2N^{1/3}}\tr\bS^2} G_{N,\beta}\left(\sqrt{2N}\I+\frac{1}{\sqrt{2N^{1/3}}}\bS\right)=\\ 
\int (d\W)\exp \left( \frac{\mathrm{i}}{3}\tr \W^3+\mathrm{i}\tr \W\bS \right)\exp\left(-\sum_{k\geq 4}\frac{\mathrm{i}^k}{kN^{(k-3)/3}}\tr \W^k\right).
\end{multline}
Clearly, the last line is similar to the matrix Airy integral plus a correction of order less or equal to $N^{-1/3}$.  However, the preceding approach is  not rigorous: it doesn't proves that Eq.\ \eqref{changeW} furnishes the {\it main contribution} to the integral when $N$ is large; it doesn't explain why $\W$ should be considered as a {\it Hermitian} $4/\beta$-matrix.  

Proposition  \ref{propEdge} only provides the dominant term to the average of products of characteristic polynomials, or equivalently to  multivariate Hermite polynomial, evaluated at the edge.  One can nevertheless find an infinite (but not convergent) asymptotic series by exploiting Eq.~\eqref{EqHeuristic}.  Note that the latter equation remains true when considering the steepest descent method used in Appendix B.   Recall the well known identity  (see e.g. \cite[Chapter I]{Mac})
\begin{equation}
 \exp\left({\sum_{n\geq1}\frac{1}{n}v_n}\right)=\sum_\lambda \frac{1}{z_\lambda}v_\lambda,
\end{equation}
where the sum is taken over all partitions $\lambda$, $ v_\lambda=v_{\lambda_1}\cdots v_{\lambda_\ell}$, and $z_\lambda$ is the quantity defined just above Eq.~\eqref{CombinatScalProd}.  The use of the latter formula in Eq.\ \eqref{EqHeuristic} leads to 
 \begin{multline}\label{EqHeuristic}
 C^{-1} e^{-N^{1/3}\tr \bS-\frac{1}{2N^{1/3}}\tr\bS^2} G_{N,\beta}\left(\sqrt{2N}\I+\frac{1}{\sqrt{2N^{1/3}}}\bS\right)=\\ 
\int (d\W)\exp \left( \frac{\mathrm{i}}{3}\tr \W^3+\mathrm{i}\tr \W\F \right)\sum_\lambda \frac{\mathrm{i}^{3\ell(\lambda)-|\lambda|}}{N^{|\lambda|/3}}\frac{1}{z_{\lambda_+}}p_{\lambda_+}(\W).
\end{multline}
In the last equation, $\lambda_+=(\lambda_1+3,\ldots,\lambda_\ell+3)$ and $p_{\lambda}(\W)=\tr \W^{\lambda_1}\cdots\tr\W^{\lambda_\ell}$.  Now,
let $\mathbf{T}$ stand for an arbitrary $n\times n$ matrix with no symmetry property (i.e., $n^2$ independent elements).  Then
\begin{equation}
 p_k(\partial_\mathbf{T})e^{\mathrm{i}\tr \W\mathbf{T}}:=\tr (\partial_\mathbf{T})^k e^{\mathrm{i}\tr \W\mathbf{T}}= \tr (\mathrm{i}\mathbf{W})^ke^{\mathrm{i}\tr \W\mathbf{T}}
\end{equation}
and the next proposition follows.

\begin{proposition}
 With the above notation, one has formally
\begin{multline}
C^{-1} e^{-N^{1/3}\tr \bS-\frac{1}{2N^{1/3}}\tr\bS^2}\,G_{N,\beta}\left(\sqrt{2N}\I+\frac{1}{\sqrt{2N^{1/3}}}\bS\right)=\\
\left[\sum_{\lambda}\frac{(-1)^{|\lambda|}}{N^{|\lambda|/3}}\frac{1}{z_{\lambda_+}}\,p_{\lambda_+}(\partial_{\mathbf{T}})\,\mathrm{Ai}^{(\beta/2)}(\mathbf{T})\right]_{\mathbf{T}=\bS}.
\end{multline}
\end{proposition}

\section{Conclusion}

New dualities between different ensembles of random Hermitian matrices  have been obtained in the article.  Their general form, valid for all $\beta>0$, has been given in Propositions~\ref{theo1}, \ref{theo2}, \ref{prop3}, and \ref{prop4}.

In Section 5.2, Proposition \ref{prop3} has been used for proving the equivalence of the formal one-matrix model and the Gaussian model with an external field.   But the comparison of Eqs.~\eqref{eqHermiteMat} and \eqref{eqFormalProd} leads to another surprising conclusion: the partition function \eqref{FormalPart} is a multivariate Hermite polynomial!  Explicitly,
\begin{equation}\label{eqZHermite}
(\prod_{i\geq 1}s_i)^N\, Z_{n,\beta}(t_1,t_2,\ldots)=\mathcal{H}^{(\beta/2)}_{(N,N,\ldots)}(s_1,s_2,\ldots)
\end{equation}
when \begin{equation}
 t_k=2\delta_{k,2}+\left(\pm \frac{2}{\beta}\right)^{k/2} p_{k}(s_1^{-1},s_2^{-1},\ldots)
\end{equation}
for all $k\geq1$.  Similarly, the partition function, for $\beta$-matrices that are both Hermitian and positive (chiral ensembles), is equivalent to a multivariate Laguerre polynomial whose partition is rectangular.  Since the partition function of a matrix model is related to the enumeration of maps, the above result suggests that there exists a combinatorial interpretation for the multivariate Hermite and Laguerre polynomials.  This would generalise Viennot's combinatorial work on classical polynomials in one variable \cite{Viennot}.

Formal matrix models have been meanly studied when $\beta=2$.  This case is simpler and many tools have been developed for calculating the ``large N'' or topological expansion of the partition functions.  A new and very general approach, which seems relevant to the question of dualities, has been introduced recently by Eynard and Orantin \cite{EynardOrantin}.  These authors have shown that the ``free energy'' (i.e., $-\ln Z$), for many matrix models or even algebraic curves, is invariant under a certain class of transformations.  It would be interesting, on the one hand, to check if these transformations include precisely the source--external field exchange considered in Section 4 and, on the other hand, to determine if the invariance of the free energy can be generalised to the $\beta\neq2$ cases.

For $\beta=1,2,4$, the large $N$ limit of the partition function \eqref{eqZHermite} (equivalently, the expectation value of products of characteristic polynomials)  has been evaluated in Section 5.3.   It is proportional to a matrix Airy (or Kontsevich) integral. The proof given  in Appendix B is valid only when one works with the matrices themselves, but not with their eigenvalues. However, the fact that the asymptotic formula has the same form for three distinct values of $\beta$ is non trivial.  Thus, it is reasonable to surmise that formula Eq.\ \eqref{eqPropEdge} remains the same for all $\beta>0$ if one uses Eq.~\eqref{generalAiry}, for the definition of the multivariate Airy function, instead of Eq.\ \eqref{matrixAiry}.  The proof is still missing.

Finally, it is remarkable that the dualities concerning the product of {\it inverse} characteristic polynomials don't involve a change in $\beta$ (for instance, compare Propositions \ref{prop1} and \ref{prop2}).  This fact certainly indicates that averages involving ratios of characteristic polynomials are much more difficult to calculate.  Recall that the knowledge of the latter quantities gives access to the eigenvalue correlation functions (or marginal densities) which, from a physical or probabilistic point of point, are of prime importance.  Note that the average of ratios of characteristic polynomials is still an open problem  for the Circular $\beta$-Ensembles \cite{Matsumoto} despite the fact that calculations in the latter ensembles are easier than in the Gaussian $\beta$-Ensembles.  A simple exercise shows, however, that the average of ratios characteristic polynomials satisfies a Calogero equation related to a superalgebra,  $gl(p|q)$ say, if  $beta$ is rational.     Superalgebraic systems of the circular (or trigonometric) type have been previously studied  by Sergeev an Veselov \cite{VS,VS2}.  The solutions to these models are supersymmetric polynomials which belong to a special family of symmetric functions in two sets of variables (see examples 23 and 24 of Chapter~I-3 in \cite{Mac}).   Preliminary calculations indicate that solutions also exist for systems of the rational type (i.e.,  Calogero models connected to Gaussian $\beta$-Ensembles).  The relation between the superalgebraic Calogero models and the eigenvalue correlation functions for the $\beta$-Ensembles will be the subject of a forthcoming paper.  

\begin{acknow}

The author is grateful to Peter J.\ Forrester for helpful discussions.  Thanks also  to Michel Berg\`ere and Bertrand Eynard
 for stimulating discussions on related subjects.    A small part of this work, supported by NSERC, was done while
  visiting the
  Centre de recherches math\'ematiques (CRM) de l'Universit\'e
 de Montr\'eal and the D\'epartement de physique de l'Universi\'e
 Laval; the author wishes to thank  John Harnad and Pierre Mathieu
 for their hospitality. 
\end{acknow}

\appendix

\section{$\beta$-Matrices}

A $N\times N$  matrix whose entries are real ($\beta=1$), complex ($\beta=2$), or quaternionic real ($\beta=4$), can be represented as
\begin{equation}
 \mathbf{X}=\mathbf{X}^0+\sum_{k=1}^{\beta-1}\mathbf{X}^k\mathbf{e}_k,
\end{equation}
 where $\X^k$ ( for $0\leq k\leq \beta-1$)  is a $N\times N$ matrix with real elements.  The ``imaginary numbers'' $\mathbf{e}_k$ satisfy $\mathbf{e}_k^2=-1$, $\mathbf{e}_i\mathbf{e}_j=-\mathbf{e}_j\mathbf{e}_i$, and $\mathbf{e}_1\mathbf{e}_2=\mathbf{e}_3$.  The conjugation is defined by
\begin{equation}
\overline{\mathbf{X}}=\mathbf{X}^0-\sum_{k=1}^{\beta-1}\mathbf{X}^k\mathbf{e}_k.
\end{equation}
$\mathbf{X}$ is Hermitian if $\mathbf{X}^\dagger:=(\overline{\mathbf{X}})^\trans=\mathbf{X}$.  Equivalently, $\X$ is Hermitian if 
\begin{equation}
(\mathbf{X^0})^\trans=\mathbf{X^0},\quad ({\mathbf{X}^k})^\trans=-\mathbf{X}^k \quad k\geq1.
\end{equation}
The case $\beta=0$ is trivial: it is obtained when $\mathbf{X}=\mathbf{X}^0=\mathrm{diag}(x_1,\ldots,x_N)$. 

Note that the case  $\beta=4$ is special since quaternions are not commutative.  Thus, in general, $\sum_{i,j} X_{i,j}Y_{j,i}\neq \sum_{i,j}Y_{i,j}X_{j,i}$ for $\beta=4$.  The correct definition of the trace for quaternionic matrices is the following:
\begin{equation}
\tr \X:=\frac{1}{2}\sum_{i=1}^N(X_{i,i}+\overline{X_{i,i}})\qquad (\beta=4).
\end{equation}
This trace is real by definition and satisfies $\tr \X\Y=\tr\Y\X$ for all quaternionic matrices $\X$ and $\Y$.  Alternatively, one can use the representation of quaternions in terms of the Pauli matrices:
 \begin{equation}
 \tr \X=\frac{1}{2}\tr P(\X) \qquad (\beta=4)
 \end{equation}
 where 
 \begin{equation}
 P(\X)=\Big[P(X_{i,j})\Big]_{i,j=1}^N,\qquad P(X_{i,j})=\left[
                                                  \begin{array}{cc}
                                                    \phantom{-}X_{i,j}^0+\mathrm{i} X_{i,j}^1 & X_{i,j}^2+\mathrm{i} X_{i,j}^3  \\
                                                    -X_{i,j}^2+\mathrm{i} X_{i,j}^3  & X_{i,j}^0-\mathrm{i} X_{i,j}^1  \\
                                                  \end{array}
                                                \right]
 \end{equation}
In other words, $P(\X)$ is $N\times N$ matrix with $2\times 2$ entries.  For all $\beta$ and $\X$ Hermitian, 
\begin{equation}\det \mathbf{X}=\exp (\tr \ln \mathbf{X})
\end{equation}
is equal to the product of the eigenvalues of $\X$.  Note however that, for $\beta=4$, $\tr P(\X)=2\tr \X$ and   
 $\det\mathbf{X}=\sqrt{\det P(\mathbf{X})}$.

Finally,  the measure on the space of Hermitian $\beta$-matrices is simply the product of the real independent elements of $\X$:
\begin{equation}
 (d\mathbf{X})=\prod_{i\leq j} d{X}^0_{i,j}  \prod_{i< j}\prod_{k=1}^{\beta-1}dX^k_{i,j}.
\end{equation}By using  $\int_{\mathbb{R}} dx e^{-ax^2+kx}=\sqrt{\pi/a}e^{k^2/4a}$, one easily shows that
\begin{equation}\label{eqGauss}
 \int (d\X) e^{-a\tr \X^2+\tr\X\Y}=\left(\frac{\pi}{a}\right)^{N/2}\left(\frac{\pi}{2a}\right)^{\beta N(N-1)/4}e^{\frac{1}{4a}\tr \Y^2},
\end{equation}
where both $\X$ and $\Y$ are Hermitian $\beta$-matrices of size $N$.

\section{Proof of Proposition \ref{propEdge}}

Let $\bS'$ be a fixed $n\times n$ Hermitian $\beta'=4/\beta$-matrix whose eigenvalues are close the spectrum edge of the $\GbE_N$:
\begin{equation}
s'_j=\sqrt{2N}+\frac{1}{\sqrt{2N^{1/6}}}s_j .
\end{equation}
The aim is to determine the asymptotic behaviour of 
\begin{equation}
G_{N,\beta}(\bS')= \mathcal{H}^{(\beta/2)}_{(N^n)}(s') = 
 \left\langle \prod_{i=1}^n\det\left( s'_j\I\pm\sqrt{\frac{2}{\beta}}\X\right)\right\rangle_{\X\in\GbE_N}
\end{equation}
The three cases (i.e., $\beta'=1,2,4$) must treated separately.  Only the $\beta'=1$ case will be detailed below since the method is easily adapted for $\beta'=2$ and $\beta'=4$.  Note that in the latter case, one has to use the appropriate definitions for the trace and the determinant (see Appendix A).   

The starting point  is the rescaled matrix integral \eqref{eqGMatrix}.  The integrand is  an analytic function in $n+n(n-1)/2$ real variables $Y_{i,j}$.  By using the Cauchy theorem,  integrations along the real axis can be transformed into contour integrals in the complex plane. By convention, $-\pi\leq\mathrm{arg}(Y_{i,j})< \pi$ and  the logarithm is defined on its principal branch.  In order to ensure the convergence of the integrals,   each variable $Y_{i,j}$ follows a path starting at $\infty e^{\ima \phi_s}$ and ending at $\infty e^{\ima \phi_e}$, where
\begin{equation}\label{cond1}
-\pi\leq\phi_s\leq-3\pi/4\quad \mbox{or}\quad 3\pi/4\leq\phi_s< \pi\quad\mbox{and}\quad \mbox -\pi/4\leq\phi_e\leq\pi/4 
\end{equation}
 The function $f(\Y)$ in  Eq.\ \eqref{eqGMatrix} has double saddle points, noted $\eta_{i,j}$, if the following conditions are statisfied:
\begin{equation}\label{cond2}
 \left.\frac{\partial}{\partial Y_{ij}}f(\Y)\right|_{Y=\eta}=0 ,\qquad \left.\frac{\partial^2}{\partial Y_{ij}\partial Y_{kl}}f(\Y)\right|_{Y=\eta}=0
\end{equation}
together with some nonvanishing third order derivatives at the saddle point.  Only constant solutions $\eta_{ij}$ (i.e., those that do not depend on the other $\eta_{kl}$) are suitable for the multidimensional steepest descent method. Direct calculations imply that the saddle point conditions \eqref{cond2} are equivalent to following system of algebraic equations:
\begin{equation}
 -4\eta_{ii}-4\ima+(\eta^{-1})_{ii}=0,\quad -4\eta_{ij}+(\eta^{-1})_{ij}=0,\quad -4-(\eta^{-1})_{ii}(\eta^{-1})_{ii}=0
\end{equation}
together with 
\begin{equation}
 -4-(\eta^{-1})_{ii}(\eta^{-1})_{jj}- (\eta^{-1})_{ij}(\eta^{-1})_{ij}=0, \quad (\eta^{-1})_{ik}(\eta^{-1})_{jk}=0
\end{equation}
for all $1\leq i,j,k\leq n$ and $i<j$.  Note that these equations, $\eta=(\eta_{ij})_{i,j}$ is interpreted as a matrix.  By writting the inverse in terms of cofactors, which means $(\eta^{-1})_{ij}=(\det \eta )^{-1}\mathrm{cof} \eta_{ji}$ where $\det(\eta)=\sum_{j} \eta_{ij} \mathrm{cof} \eta_{ij}$ for some $i$,  one shows that the unique set of solutions is:
\begin{equation}
 \eta_{jj}=\frac{1}{2\ima} \quad\mbox{and}\quad \eta_{ij}=0
\end{equation}
for all $1\leq i<j\leq n$.  

Now, one has to ensure that the descent of $f(\Y)$ is maximum when the variables $Y_{ij}$ approach and leave $\eta_{ij}$.  Let $\theta_{kl}$ denotes the argument of the complex variable $Y_{kl}=\eta_{kl}$  and let
\begin{equation}
 \phi_{jk}=\mathrm{arg}\left.\frac{\partial^3}{ \partial Y_{kl}^3} f(\Y)\right|_{\eta_{kl}}=-\frac{\pi}{2}
\end{equation}
for all $1\leq k\leq l\leq n$.  The contours of steepest descent must comply with
\begin{equation}
 e^{3\ima \theta_{jk}+\ima \phi_{jk}}=-1.
\end{equation}
The three possible solutions are $\theta_{kl}=-5\pi/6,-\pi/6,$ and $\pi/2$.  The two former angles are compatibles with conditions \eqref{cond1}.   Set 
\begin{equation}
 W_{kl}=2N^{1/3}(Y_{kl}-\eta_{kl})
\end{equation}
Each  variable $W_{kl}$ follows the steepest descent path $\mathcal{D}$: it starts at $\infty e^{-5\pi\ima/6}$, passes through the origin, and stops at  $\infty e^{-\pi\ima/6}$.  In order to get a simple expression for $G_{N,\beta}(\bS')$, it is convenient to define the matrix $\W$ with elements $W_{kl}=W_{lk}$.  Then, by collecting all the factors coming from $f(\Y)$ and $g(\Y,\bS)$ in Eq.\ \eqref{eqGMatrix}, one shows that $G_{N,\beta}(\bS')$ is equal to 
{\small \begin{equation}\label{limG1}
 Ce^{N^{1/3}\tr \bS+\frac{1}{2N^{1/3}}\tr\bS^2}  
\int_{\mathcal{D}} (d\W)\exp \left(- \frac{\mathrm{i}}{3}\tr \W^3-\mathrm{i}\tr \W\bS \right)\exp\left(-\sum_{k\geq 4}\frac{1}{k\mathrm{i}^kN^{(k-3)/3}}\tr \W^k\right).
\end{equation}
}where $\int_{\mathcal{D}} (d\W)$ stands for the $n+n(n-1)/2$ contour integrals along the path $\mathcal{D}$, and $C$ is the constant given in Eq.\ \eqref{eqC}.  Therefore, the following limit holds :
\begin{equation}\label{limG2}
 \lim_{N\rightarrow \infty} C^{-1}e^{-N^{1/3}\tr \bS-\frac{1}{2N^{1/3}}\tr\bS^2}G_{N,\beta}(\bS') = \int_{\mathcal{D}} (d\W)\exp \left( -\frac{\mathrm{i}}{3}\tr \W^3-\mathrm{i}\tr \W\bS \right).
\end{equation}
Obviously, the first correction to the above limit is of order less or equal to $N^{-1/3}$.  The complex functions involved in the integrals of the last equation are analytic, so Cauchy's theorem can be applied once again.  The integrals remains convergent if each variable goes from $\infty e^{i\theta_s}$ to $\infty e^{i\theta_s}$ where
\begin{equation}
-\pi\leq\theta_s\leq-2\pi/3\quad \mbox{and}\quad \mbox -\pi/3\leq\theta_e\leq 0. 
\end{equation}
 Contours along the real line are chosen so that $\W$ can be interpreted as a real symmetric matrix.  The change $\W\mapsto-\W$ finishes the proof.

\end{document}